\documentclass[journal,transmag]{IEEEtran}

\usepackage{amsmath,amsthm,amssymb}
\usepackage{amsfonts}
\usepackage{graphicx}
\usepackage{hyperref}
\usepackage[utf8]{inputenc}
\usepackage{tikz, caption}
\usepackage{multirow}
\usepackage{enumitem}

\usetikzlibrary{shapes.geometric, arrows}

\tikzstyle{startstop} = [circle, minimum width=0.1cm, minimum height=0.1cm,text centered, draw=white]
\tikzstyle{arrow} = [thick,->,>=stealth]
\tikzstyle{rectangleSolid} = [rectangle, draw, fill=white!20, node distance=2cm, text width=4em, text centered,  minimum height=2em, thick]
\tikzstyle{rectangleSolidSmall} = [rectangle, draw, node distance=2cm, text width=1em, text centered,  minimum height=1em, thick]
\tikzstyle{rectangleDashed} = [rectangle, draw, rounded corners,minimum width=7em, minimum height=12em,dashed]
\tikzstyle{rectangleNoBorders} = [rectangle, rounded corners,minimum width=1em, minimum height=1em,draw=white]
\tikzstyle{rectangleOpaque} = [rectangle, draw=white, node distance=2cm, text width=4em, text centered, rounded corners, minimum height=2em, thick]
\tikzstyle{line} = [thick,-]
\tikzstyle{romb} = [diamond, draw, text centered, inner sep=3pt]
\tikzstyle{rectangleSolidWide} = [rectangle, draw, fill=white!20, node distance=2cm, text width=4.5em, text centered,  minimum height=2em, thick]
\tikzstyle{rectangleOpaqueWide} = [rectangle, draw=white, node distance=2cm, text width=4.5em, text centered, rounded corners, minimum height=2em, thick]

\begin{document}

\title{Fingerprint template protection\\ using minutia-pair spectral representations}

\author{\IEEEauthorblockN{Taras Stanko$^1$, Bin Chen$^{1,2}$, Boris \v{S}kori\'{c}$^1$}
\IEEEauthorblockA{\small $^1$ Eindhoven University of Technology 
\quad
$^2$ School of Computer and Information, Hefei University of Technology}
}


\IEEEtitleabstractindextext{%
\begin{abstract}
Storage of biometric data requires some form of template protection
in order to preserve the privacy of people enrolled in a biometric database.
One approach is to use a Helper Data System.
Here it is necessary to transform the raw biometric measurement into
a fixed-length representation.
In this paper we extend the spectral function approach of Stanko and \v{S}kori\'{c} \cite{SS2017},
which provides such a fixed-length representation for fingerprints.
First, we introduce a new spectral function that captures different information from the
minutia orientations. It is complementary to the original spectral function, and we use both of them
to extract information from a fingerprint image.
Second, we construct a helper data system consisting of zero-leakage quantisation
followed by the Code Offset Method.
We show empirical data which demonstrates that applying our helper data system
causes only a small performance penalty compared to fingerprint authentication based on
the unprotected spectral functions.

\end{abstract}

}

\maketitle

\setlength{\parindent}{0mm}

\IEEEdisplaynontitleabstractindextext

%
\IEEEpeerreviewmaketitle

\section{Introduction}

\subsection{Biometric template protection}
\label{sec:introprot}

Biometric authentication has become popular because of its convenience. 
Biometrics cannot be forgotten or left at home. 
Although biometric data is not exactly secret (we are leaving a trail of fingerprints, DNA etc.), 
it is important to protect biometric data for privacy reasons. 
Unprotected storage of biometric data could reveal medical conditions and would allow cross-matching of entries in different databases. 
Large-scale availability of unprotected biometric data would make it easier for malevolent parties to leave misleading traces at crime scenes 
(e.g.\,artificial fingerprints \cite{matsumoto2002}, synthesized DNA \cite{FWDG2010}.) 
One of the easiest ways to properly protect a biometric database against breaches 
and insider attacks (scenarios where the attacker has access to decryption keys)
is to store biometrics in hashed form, 
just like passwords.
An error-correction step has to be added to get rid of the measurement noise. 
To prevent critical leakage from the error correction redundancy data, 
one uses a Helper Data System (HDS) \cite{LT2003}, \cite{dGSdVL}, \cite{SAS2016}, 
for instance a Fuzzy Extractor or a Secure Sketch \cite{JW99}, \cite{DORS2008}, \cite{CFPRS2016}. 
The best known and simplest HDS scheme is the code-offset method (COM). 
The COM utilizes a linear binary error-correction code and thus requires a fixed-length representation of the biometric measurement. 
Such a representation is not straightforward when the measurement noise can cause features of the biometric to appear/disappear. 
For instance, some minutiae may not be detected in every image captured from the same finger.

A fixed-length representation called {\em spectral minutiae} was introduced by Xu et al. \cite{XVBKAG2009}, \cite{XV2009}, \cite{XV2009CISP}, \cite{XuVeldhuis2010}.
For every detected minutia of sufficient quality, the method 
evaluates a Fourier-like spectral function on a fixed-size two-dimensional grid;
the contributions from the different minutiae are added up. 
Disappearance of minutiae or appearance of new ones does not affect the size of this representation.

One of the drawbacks of Xu et al.'s construction is that phase information is discarded in order to
obtain translation invariance.
Nandakumar \cite{Nanda2010} proposed a variant 
which does not discard the phase information.  
However, it reveals personalised reliability data, which makes it  
difficult to use in a privacy-preserving scheme.

A minutia-{\it pair} based variant of Xu et al.'s technique was introduced in \cite{SS2017}.
It has a more compact grid and reduced computation times.
Minutia pairs (and even triplets) were used in
\cite{FBJR2007,JTOT2010}, but with a different
attacker model that allows 
encryption keys to exist that are not accessible to the attacker.

\subsection{Contributions and outline}
\label{sec:contrib}

First we extend
the pair-based spectral minutiae method \cite{SS2017}
by introducing a new spectral function that captures different information from 
the minutia orientations.
Then we use the spectral functions
as the basis for a template protection system.
Our HDS consists of two stages.
In the first stage, we discretise the analog spectral representation using a zero-leakage
HDS \cite{dGSdVL,SAS2016}. This first HDS reduces quantisation noise, and the helper data reveals no information about
the quantised data.
Discretisation of the spectral functions typically yields only one bit per grid point. 
We concatenate the discrete data
from all the individual grid points into one long bitstring.
In the second stage we apply the Code Offset Method.
Our code of choice is a Polar code, because Polar code are low-complexity capacity-achieving codes 
with flexible rate.

We present False Accept vs.\;False Reject tradeoffs at various stages of the data processing.
We introduce the `superfinger' enrollment method, in which we average 
the spectral functions from multiple enrollment images.
By combining three enrollment images in this way, and constructing 
a polar code specifically tuned to the individual bit error rate of each
bit position, we achieve an Equal Error Rate around 1\% for a high-quality 
fingerprint database, and around 6\% for a low-quality database.

The outline of the paper is as follows.
In Section~\ref{sec:prelim} we introduce notation
briefly review helper data systems, the spectral minutiae representation,
and polar codes.
In Section~\ref{sec:newM} we introduce the new spectral function.
In Section~\ref{sec:approachexp} we explain our experimental approach
and motivate certain design choices such as the number of discretisation intervals
and the use of a Gaussian approximation.
We introduce two methods for averaging enrollment images.

Section~\ref{sec:results} contains our results, mostly in the form of ROC curves.
In Section~\ref{sec:discussion} we discuss the results and identify 
topics for future work.


\section{Preliminaries} 
\label{sec:prelim}

\subsection{Notation and terminology}
We use capitals to represent random variables, and lowercase for their realizations. 
Sets are denoted by calligraphic font. The set $\mathcal{S}$ is defined as $\mathcal{S}=\{0,\ldots,N-1\}$.
The mutual information (see e.g.\,\cite{CoverThomas}) between $X$ and $Y$ is $I(X;Y)$.
The probability density function (pdf) of the random variable $X\in\mathbb{R}$ in written as 
$f(x)$ and its cumulative distribution function (cdf) as~$F(x)$.
We denote the number of minutiae found in a fingerprint by~$Z$.
The coordinates of the $j$'th minutia are ${\boldsymbol x}_j=(x_j,y_j)$ and its orientation
is $\theta_j$.
We write $\boldsymbol x=(\boldsymbol x_j)_{j=1}^Z$ and 
$\boldsymbol\theta=(\boldsymbol \theta_j)_{j=1}^Z$
We will use the abbreviations FRR = False Reject Rate,  
FAR = False Accept Rate, EER = Equal Error Rate, ROC = Receiver Operating Characteristic.
Bitwise xor of binary strings is denoted as~$\oplus$.

\subsection{Helper Data Systems}
\label{sec:prelimHDS}

\begin{figure}[h]
	\centerline{\includegraphics[width=45mm]{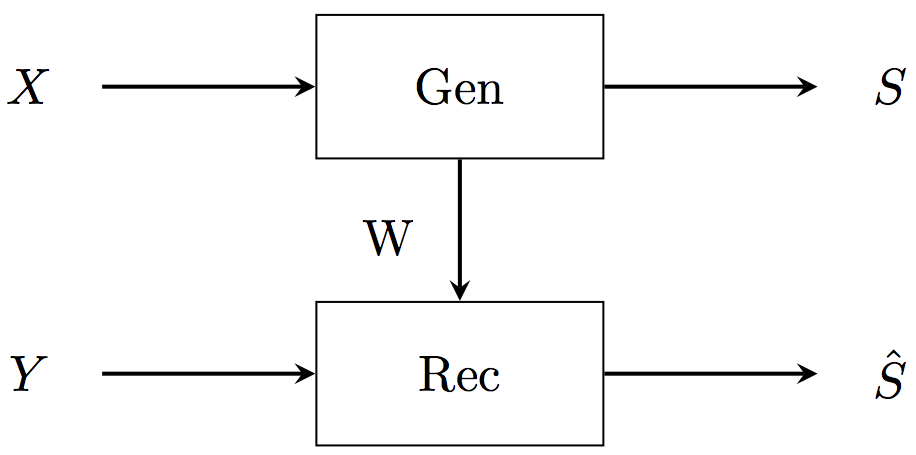}}
	\caption{\it Data flow in a generic Helper Data System.}
	\label{fig:HDS}
\end{figure}

A HDS is a cryptographic primitive that allows one to reproducibly
extract a secret from a noisy measurement. 
A HDS consist of two algorithms: {\tt Gen} (generation) and {\tt Rec} (reconstruction), see Fig. \ref{fig:HDS}. 
The {\tt Gen} algorithm takes a measurement $X$ as input and generates the secret $S$ and a helper data~$W$. 
The {\tt Rec} algorithm has as input a noisy measurement $Y$ and the helper data; it outputs an estimator $\hat S$.
If $Y$ is sufficiently close to $X$ then $\hat S=S$. 
The helper data should not reveal much about $S$. Ideally it holds that $I(W;S)=0$. This is known as {\em Zero Leakage}
helper data.

\subsection{Two-stage HDS template protection scheme}
\label{sec:prelim2stage}

Fig.\,\ref{fig:twostageHDS} shows the two-stage HDS architecture
mentioned in Section~\ref{sec:contrib}.
The enrollment measurement $x$ is transformed to the spectral representation $(x_i)_{i=1}^M$
on $M$ grid points. 
The first-stage enrollment procedure
{\tt Gen1} is applied to each $x_i$ individually, yielding
short (mostly one-bit) secrets $s_i$ and zero-leakage helper data $w_i$.
The $s_1\ldots s_M$ are concatentated into a string~$k$. 
Residual noise in $k$
is dealt with by the second-stage HDS (Code Offset Method), whose {\tt Gen2} produces
a secret $c$ and helper data~$r$.
A hash $h(c||z)$ is computed, where $z$ is salt. 
The hash and the salt are stored.

In the verification phase, the noisy $y$ is processed as shown in the bottom half of
Fig.\,\ref{fig:twostageHDS}. 
The reconstructed secret $\hat c$ is hashed with the salt $z$; the resulting hash is compared to
the stored hash.


\begin{center}
\begin{figure*}

	\begin{tikzpicture}[node distance=4cm]

\node[rectangleNoBorders,xshift=2.9 cm, yshift=+0.3 cm](x1){$x_1$};
\node[rectangleNoBorders,xshift=2.9 cm, yshift=-0.3 cm](x1){$x_M$};
\node[rectangleNoBorders,xshift=2.9 cm, yshift=-2.2 cm](x1){$y_1$};
\node[rectangleNoBorders,xshift=2.9 cm, yshift=-2.8 cm](x1){$y_M$};
\node[rectangleNoBorders,xshift=3.9 cm, yshift=-0.8 cm](x1){$w_1$};
\node[rectangleNoBorders,xshift=4.9 cm, yshift=-0.8 cm](x1){$w_M$};
\node[rectangleNoBorders,xshift=6.0 cm, yshift=+0.3 cm](x1){$s_1$};
\node[rectangleNoBorders,xshift=6.0 cm, yshift=-0.3 cm](x1){$s_M$};
\node[rectangleNoBorders,xshift=6.0 cm, yshift=-2.2 cm](x1){$\hat{s}_1$};
\node[rectangleNoBorders,xshift=6.0 cm, yshift=-2.8 cm](x1){$\hat{s}_M$};
\node[rectangleNoBorders,xshift=4.5 cm, yshift=+1.0cm](x1){Stage 1 HDS};
\node[rectangleNoBorders,xshift=10.0 cm, yshift=1.0 cm](x1){Stage 2 HDS};
\node[rectangleNoBorders,xshift=10.25 cm, yshift=-0.8 cm](x1){$r$};
\node[rectangleNoBorders,xshift=12.75 cm, yshift=0.25 cm](x1){$h(c||z)$};
\node[rectangleNoBorders,xshift=12.75 cm, yshift=-2.25 cm](x1){$h(\hat{c}||z)$};

\draw[line] (11.5,-0.8) -- (11.9,-0.8)node [above,pos=0.06] {$z$}   ;
\node  [startstop] (x) {};
\node [rectangleOpaque, right of=x, xshift=-0.25cm, yshift=0.15 cm] (transform3){};
\node [rectangleOpaque, right of=transform3, xshift=-2.0 cm, yshift=-0.3 cm] (transform2){};
\node [rectangleSolid, right of=transform2, xshift=-2.0 cm,yshift=+0.15 cm] (transform1) {transform};
\node [rectangleSolid, right of=transform3, xshift=0.8 cm](Gen1upup) {};
\draw [arrow]  (transform3) -- node [anchor=south]{}(Gen1upup);
\node [rectangleSolid, right of=transform1, xshift=0.65 cm](Gen1up) {};
\draw [arrow]  (transform1) -- node [anchor=south]{}(Gen1up);
\node [rectangleSolid, right of=transform2, xshift=0.5 cm](Gen1) {Gen 1};
\draw [arrow]  (transform2) -- node [anchor=north]{}(Gen1);
\node [rectangleOpaqueWide, right of=Gen1upup, xshift=0.85 cm](concatenate3) {};
\draw [arrow]  (Gen1upup) -- node [anchor=south]{}(concatenate3);
\node [rectangleOpaqueWide, right of=Gen1, xshift=1.15 cm](concatenate1) {};
\draw [arrow]  (Gen1) -- node [anchor=north]{}(concatenate1);
\node [rectangleSolidWide, right of=Gen1up, xshift=1.0 cm](concatenate2) {concatenate};
\draw [arrow]  (Gen1up) -- node [anchor=south]{}(concatenate2);
\node [rectangleSolid, right of=concatenate2, xshift=0.5cm](Gen2) {Gen 2};
\node [rectangleSolidSmall, right of=Gen2, xshift=-0.0cm](h) {$h$};
\draw [arrow]  (x) -- node [anchor=south east]{x}(transform1);
\draw [arrow]  (concatenate2) -- node [anchor=south east]{$k$}(Gen2);
\draw [arrow]  (Gen2) -- node [anchor=south west]{$c$}(h);

\node  [startstop, yshift=-2.5 cm] (y) {};
\node [rectangleOpaque, right of=y, xshift=-0.25cm, yshift=0.15 cm] (transform6){};
\node [rectangleOpaque, right of=y, xshift=-0.25cm, yshift=-0.15 cm] (transform4){};
\node [rectangleSolid, right of=y, xshift=-0.25 cm, yshift=0 cm] (transform5) {transform};
\node [rectangleSolid, right of=transform6, xshift=0.8 cm, yshift=0 cm](Rep1upup) {};
\draw [arrow]  (transform6) -- node [anchor=south east]{}(Rep1upup);
\node [rectangleSolid, right of=transform5, xshift=0.65 cm, yshift=0 cm](Rep1) {};
\draw [arrow]  (transform5) -- node [anchor=north]{}(Rep1);
\node [rectangleSolid, right of=transform4, xshift=0.5 cm, yshift=0 cm](Rep1up) {Rep 1};
\draw [arrow]  (transform4) -- node [anchor=north east]{}(Rep1up);
\draw [arrow]  (y) -- node [anchor=south east]{y}(transform5);
\node [rectangleOpaqueWide, right of=Rep1, xshift=1.0 cm, yshift=0.15 cm](concatenate6) {};
\node [rectangleOpaqueWide, right of=Rep1, xshift=1.0 cm, yshift=-0.15 cm](concatenate5) {};
\node [rectangleSolidWide, right of=Rep1, xshift=1.0 cm](concatenate4) {concatenate};
\draw [arrow]  (Rep1) -- node [anchor=north]{}(concatenate4);
\draw [arrow]  (Rep1up) -- node [anchor=north west]{}(concatenate5);
\draw [arrow]  (Rep1upup) -- node [anchor=south west ]{}(concatenate6);
\node [rectangleSolid, right of=concatenate4, xshift=0.5cm](Rep2) {Rep 2};
\node [rectangleSolidSmall, right of=Rep2, xshift=0.0cm](h2) {h};
\node[romb, right of = h2, xshift = -2.25cm](romb){$_{=}^{?}$};
\draw [arrow]  (concatenate4) -- node [anchor=south east]{$\hat{k}$}(Rep2);
\draw [arrow]  (Rep2) -- node [anchor=south west]{$\hat{c}$}(h2);
\draw [arrow]  (h2) -- node [anchor=south]{}(romb);


\draw [arrow]  (Gen1up) -- node [anchor=north east ]{}(Rep1);
\draw [arrow]  (Gen1) -- node [anchor=east]{}(Rep1up);
\draw [arrow]  (Gen1upup) -- node [anchor=west]{}(Rep1upup);
\draw [arrow]  (Gen2) -- node [anchor=east]{}(Rep2);
\draw[<->] (h) -- (h2);
\draw [arrow]  (h) -| node [anchor=south east]{}(romb);
\draw[line] (0,-1.25) -- (15,-1.25)node [above,pos=0.06] {Enrollment} node [below,pos=0.06	] {Verification}  ;
\node[rectangleDashed,xshift=4.4 cm, yshift=-1.25 cm](brr1){};
\node[rectangleDashed,xshift=9.85 cm, yshift=-1.25 cm](brr2){};
\node [rectangleOpaque, right of = romb, xshift = -0.25cm](Accept){};
\draw [arrow]  (romb) -- node [anchor=south ]{}(Accept);
\node [rectangleSolid, right of=transform2, xshift=0.5 cm](Gen1) {Gen 1};	
\node[text width=3cm] at (16.25,-2.5) {Accept/Reject};
\end{tikzpicture}
\caption{\it Two-stage Helper Data System.}
\label{fig:twostageHDS}
\end{figure*}
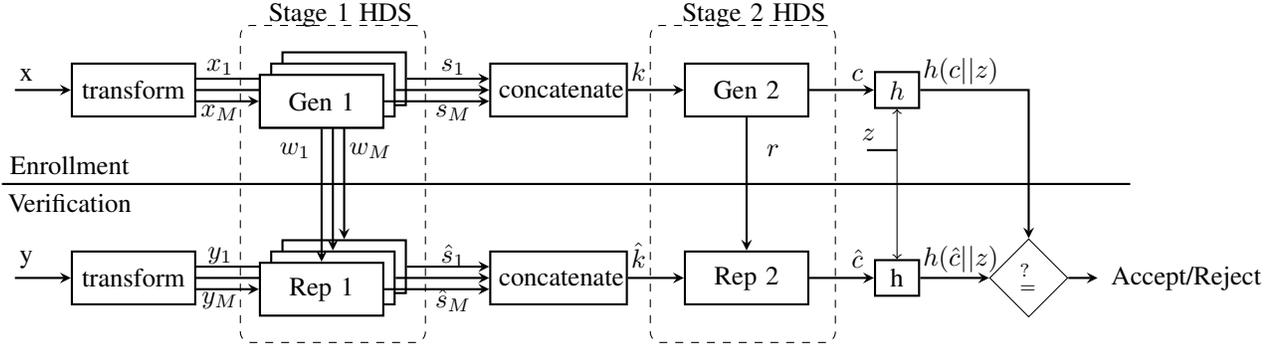
\end{center}

\subsection{Minutia-pair spectral representation} 
\label{sec:prelimspectral}

Minutiae are special features in a fingerprint, e.g. ridge endings and bifurcations. 
We briefly describe the minutia-pair spectral representation introduced in \cite{SS2017}.
For minutia indices $a,b\in\{1,\ldots,Z\}$
the distance and angle between these minutiae are given by
$R_{ab}=|{\bf x}_a-{\bf x}_b|$
and 
$\tan \phi_{ab}= \frac{y_a-y_b}{x_a-x_b}$.
The spectral function $\cal M_{\boldsymbol{x\theta}}$ is defined as
\begin{equation}
\label{defMxTheta}
	\mathcal{M}_{\boldsymbol{x\theta}}(q,R) = 
	\sum_{a=1}^{Z}  \sum_{b=a+1}^{Z} e^{iq\phi_{ab}}  e^{-\frac{(R-R_{ab})^2}{2\sigma^2}} e^{i (\theta_b - \theta_a)},
\end{equation} 
where $\sigma$ is a width parameter.
The spectral function is evaluated on a discrete $(q,R)$ grid.
The variable $q$ is integer and can be interpreted as the Fourier conjugate of an angular variable,
i.e.\,a harmonic. 
The function $\cal M_{\boldsymbol{x\theta}}$ is invariant under translations of $\boldsymbol x$.
When a rotation of the whole fingerprint image is applied over an angle $\delta$,
the spectral function transforms in a simple way,
\begin{equation}
	\mathcal{M}_{\boldsymbol{x\theta}}(q,R) \to e^{iq\delta} \mathcal{M}_{\boldsymbol{x\theta}}(q,R).
\label{Mrot}
\end{equation}

\subsection{Zero Leakage Helper Data Systems}
\label{sec:ZLHDS}

We briefly review the ZLHDS developed in \cite{dGSdVL,SAS2016}
for quantisation of an enrollment measurement $X\in\mathbb R$.
The density function of $X$ is $f$, and the cumulative distribution function is~$F$.
The verification measurement is~$Y$.
The $X$ and $Y$ are considered to be noisy versions of an underlying `true' value.
They have zero mean and variance $\sigma_X^2$, $\sigma_Y^2$, respectively. 
The correlation between $X$ and $Y$ can be characterised by writing 
$Y=\lambda X+V$, where $\lambda\in[0,1]$ is the attenuation parameter and $V$
is zero-mean noise independent of $X$, with variance $\sigma_V^2$.
It holds that $\sigma^2_Y =\lambda \sigma^2_X+\sigma^2_V$. 
We consider the {\em identical conditions} case: 
the amount of noise is the same during enrollment and reconstruction.
In this situation we have $\sigma^2_X = \sigma^2_Y$ and $\lambda^2 = 1-\frac{\sigma^2_V}{\sigma^2_X}$.

The real axis $\mathbb R$ is divided into $N$ intervals ${\cal A}_\alpha=(\Omega_{\alpha},\Omega_{\alpha+1})$, 
with $\alpha\in\cal S$,
${\cal S}=\{0,\ldots,N-1\}$.
Let $p_\alpha=\Pr[X\in{\cal A}_\alpha]$.
The quantisation boundaries are given by
$\Omega_\alpha=F^{\rm inv}(\sum_{j=0}^{\alpha-1}p_j)$.
The {\tt Gen} algorithm produces the secret $s$ as $s={\rm max} \{ \alpha \in \mathcal{S}: x\geq \Omega_\alpha   \}$
and the helper data $w\in[0,1)$
as $w=[F(x)-\sum_{j=0}^{s-1}p_j]/p_s$.
The inverse relation, for computing $x$ as a function of $s$ and $w$, is given by
$\xi_{s,w}=F^{\rm inv}(\sum_{j=0}^{s-1}p_j+wp_s)$.

The {\tt Rec} algorithm computes the estimator $\hat s$ as the value in $\cal S$
for which it holds that
$y\in(\tau_{\hat s,w}, \tau_{\hat s+1,w})$,
where the parameters $\tau$ are decision boundaries.
In the case of Gaussian noise these boundaries are given by
\begin{equation}
	\tau_{\alpha,w}=\lambda\frac{\xi_{\alpha-1,w}+\xi_{\alpha,w}}2
	+\frac{\sigma_V^2 \ln\frac{p_{\alpha-1}}{p_\alpha}}
	{\lambda(\xi_{\alpha,w}-\xi_{\alpha-1,w})}.
\end{equation}
Here it is understood that $\xi_{-1,w}=-\infty $ and $\xi_{N,w}=\infty$, 
resulting in $\tau_{0,w}=-\infty$, $\tau_{N,w}=\infty$.

The above scheme ensures that $I(W;S)=0$ and that the reconstruction 
errors are minimized.

\subsection{The Code Offset Method (COM)}
\label{sec:prelimCOM}

We briefly describe how the COM is used as a Secure Sketch.
Let $C$ be a linear binary error correcting code with message space $\{0, 1\}^m$ and codewords in $\{0, 1\}^n$. 
It has an encoding ${\tt Enc}$: $\{0, 1\}^m \to \{0, 1\}^n$,
a syndrome function $\tt Syn$: $\{0,1\}^n \to \{0,1\}^{n-m}$
and a syndrome decoder $\tt SynDec$: $\{0, 1\}^{n-m} \to \{0, 1\}^n$.
In Fig.\,\ref{fig:twostageHDS} the {\tt Gen2} computes
the helper data $r$ as $r={\tt Syn}\,k$.
The $c$ in Fig.\,\ref{fig:twostageHDS} is equal to~$k$.
The {\tt Rep2} computes the reconstruction 
$\hat k=k'\oplus{\tt SynDec}(r\oplus{\tt Syn}\,k')$.


\subsection{Polar codes}
\label{sec:prelimPolar}

Polar codes, proposed by Ar{\i}kan \cite{Arikan2009},
are a class of linear block codes that 
get close to the Shannon limit even at small code length.
They are based on the repeated application of the {\em polarisation} operation
{\small{$\Big(\begin{matrix}
1 &0\\
1 &1
\end{matrix}\Big)$}}
on two bits of channel input.
Applying this operation creates two virtual channels, one of which is better than the original channel
and one worse.
For $n$ channel inputs, repeating this procedure in the end yields $m$ near-perfect virtual channels,
with $m/n$ close to capacity, and $n-m$ near-useless channels.
The $m$-bit message is sent over the good channels, while the bad ones are `frozen', i.e  used to send a fixed string
known a priori by the recipient.

The most popular decoder is the Successive Cancellation Decoder (SCD),
which sequentially estimates 
message bits $(c_i)_{i=1}^m$ according to the frozen bits and the previously estimated 
bits $\hat c_{i-1}$.
Polar codes have been recently adopted for the next generation wireless standard (5G),
especially for control channels,
which have short block length ($\leq1024$).



\section{A new spectral function}
\label{sec:newM}

\begin{figure}[b]
\begin{center}
\includegraphics[scale=0.45]{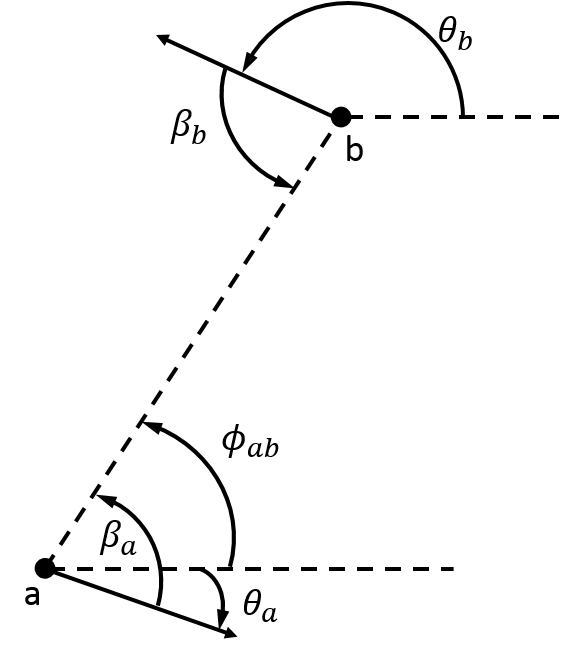}
\caption{\it The relevant angles in a minutia pair.} 
\label{fig:anglesBeta}
\end{center}
\end{figure}


Consider Fig.\,\ref{fig:anglesBeta} (modified from \cite{BozorthNIST}).
The invariant angle $\beta_a$ is defined as the angle from the orientation of minutia $a$ to the connecting line $ab$,
taken in the positive direction.
(The $\beta_b$ is defined analogously).
Modulo $2\pi$
it holds that $\theta_a+\beta_a=\phi_{ab}$
and $\theta_b+\beta_b=\phi_{ab}+\pi$.
The spectral function (\ref{defMxTheta}) uses only the invariant angle
$\beta_a-\beta_b+\pi=\theta_b-\theta_a$.
The second invariant angle, which can be written e.g. as
$\pi-\beta_a-\beta_b=\theta_a+\theta_b-2\phi_{ab}$, is not used.
We therefore now introduce a new spectral function, denoted as $\mathcal{M}_{x\beta}$,
which incorporates the invariant angle $\pi-\beta_a-\beta_b$.

\begin{equation}
\label{defMxBeta}
\mathcal{M}_{\boldsymbol{x\beta}}(q,R) = 
\sum_{a=1}^{Z}  \sum_{b=a+1}^{Z} e^{i\phi_{ab}(q-2)}  e^{-\frac{(R-R_{ab})^2}{2\sigma^2}} e^{i (\theta_b + \theta_a)}.
\end{equation}

We will use $\mathcal{M}_{x\theta}$, $\mathcal{M}_{x\beta}$ and their fusion.
\section{Experimental approach}
\label{sec:approachexp}

\subsection{Databases}
We use the MCYT, FVC2000, and FVC2002 database. 
The MCYT database \cite{MCYT} contains good-quality images from 100 individuals: 10 fingers per individual and 12 images per finger. 
FVC2000 and FVC2002 contain low-quality images (only index and middle fingers \cite{FVCs}). 
Each FVC database contains 100 fingers, 8 images per finger. 
In FVC2002, images number 3, 4, 5, and 6 have an exceptionally large 
angular displacement, so they are omitted from the experiments. 

We extract the minutia position and orientation $(x_j,y_j,\theta_j)$ by using VeriFinger software \cite{VeriFinger}. 
For MCYT we evaluate the spectral functions on the same grid as \cite{SS2017}, namely
$R\in\{16,22,28,\ldots,130\}$ and $q \in\{1,2,\ldots,16\}$ 
and we maintain $\sigma=2.3$ pixels. 
For the FVC databases we use the same grid, and $\sigma=3.2$ pixels turns
out to be a good choice.
The average number of minutiae that can be reliably found is $Z=35$.

\subsection{No image rotation}

As mentioned in \cite{SS2017}, during the reconstruction procedure one can try
different rotations of the verification image, but it results only in a minor
improvement of the EER.
For this reason we do not apply image rotation.

\subsection{Quantization methods}
Before quantization all spectral functions are normalized to zero mean and unit variance,
where the variance is taken of the real and imaginary part together.
We quantize the real and imaginary part of the spectral functions separately. 
We study two methods: `hard thresholding' (without helper data) and the Zero Leakage
quantisation of Section~\ref{sec:prelimHDS}.
The hard thresholding gives a bit value `1' if ${\rm Re}\, M>0$ and `0' otherwise.
We will show results for this method mainly to demonstrate the advantages
of Zero Leakage quantisation.

\subsection{Gaussian probability distributions}
\label{sec:gaussian}

\begin{figure}[h]
	\centerline{\includegraphics[width=90mm]{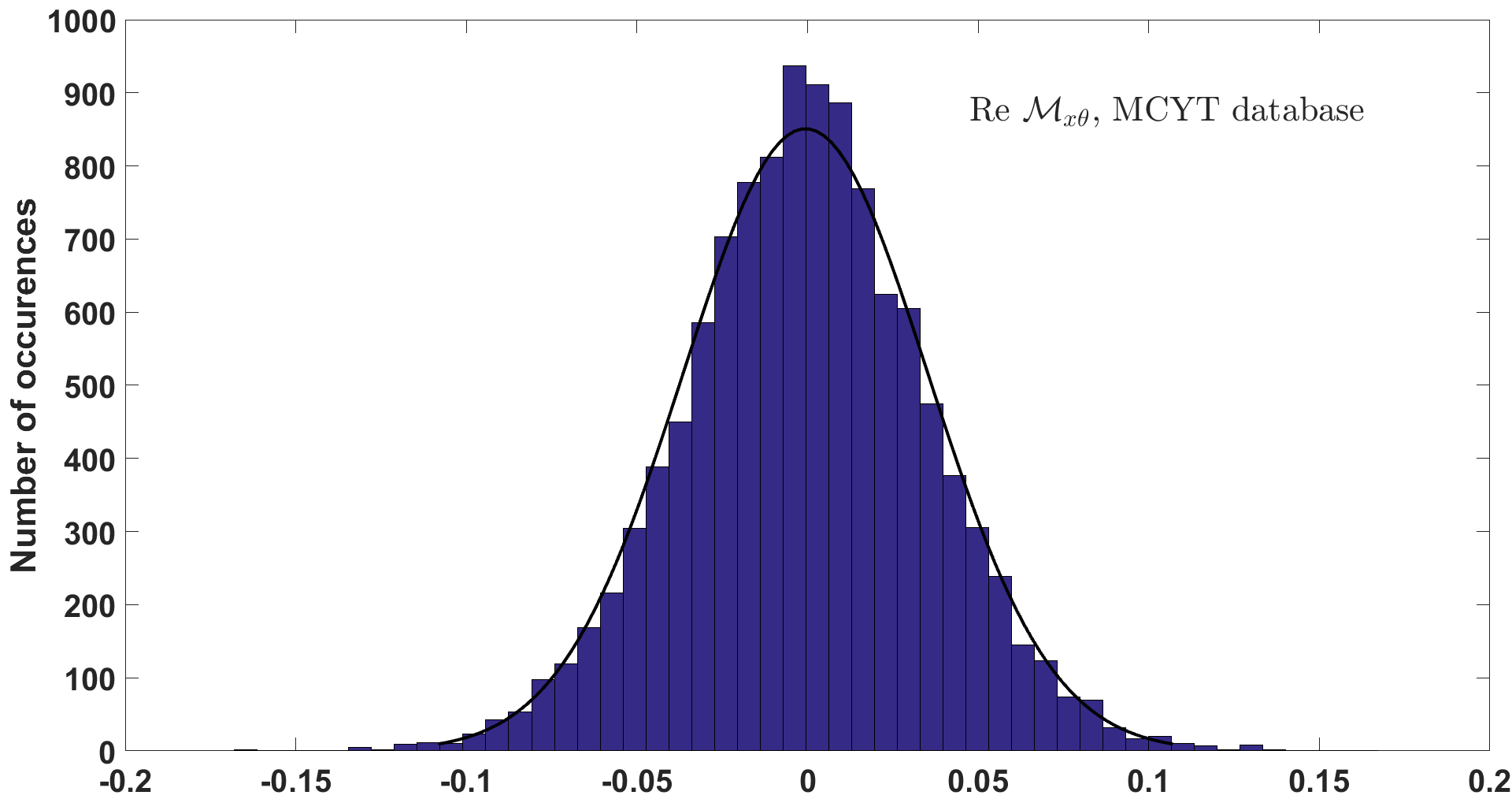}}
\caption{\it Histogram of ${\rm Re}\,\mathcal{M}_{x\theta}$, and a fitted Gaussian.}
\label{fig:HistogramValueOfMqthetaRealMCYT}
\end{figure}

\begin{figure}[h]
	\centerline{\includegraphics[width=90mm]{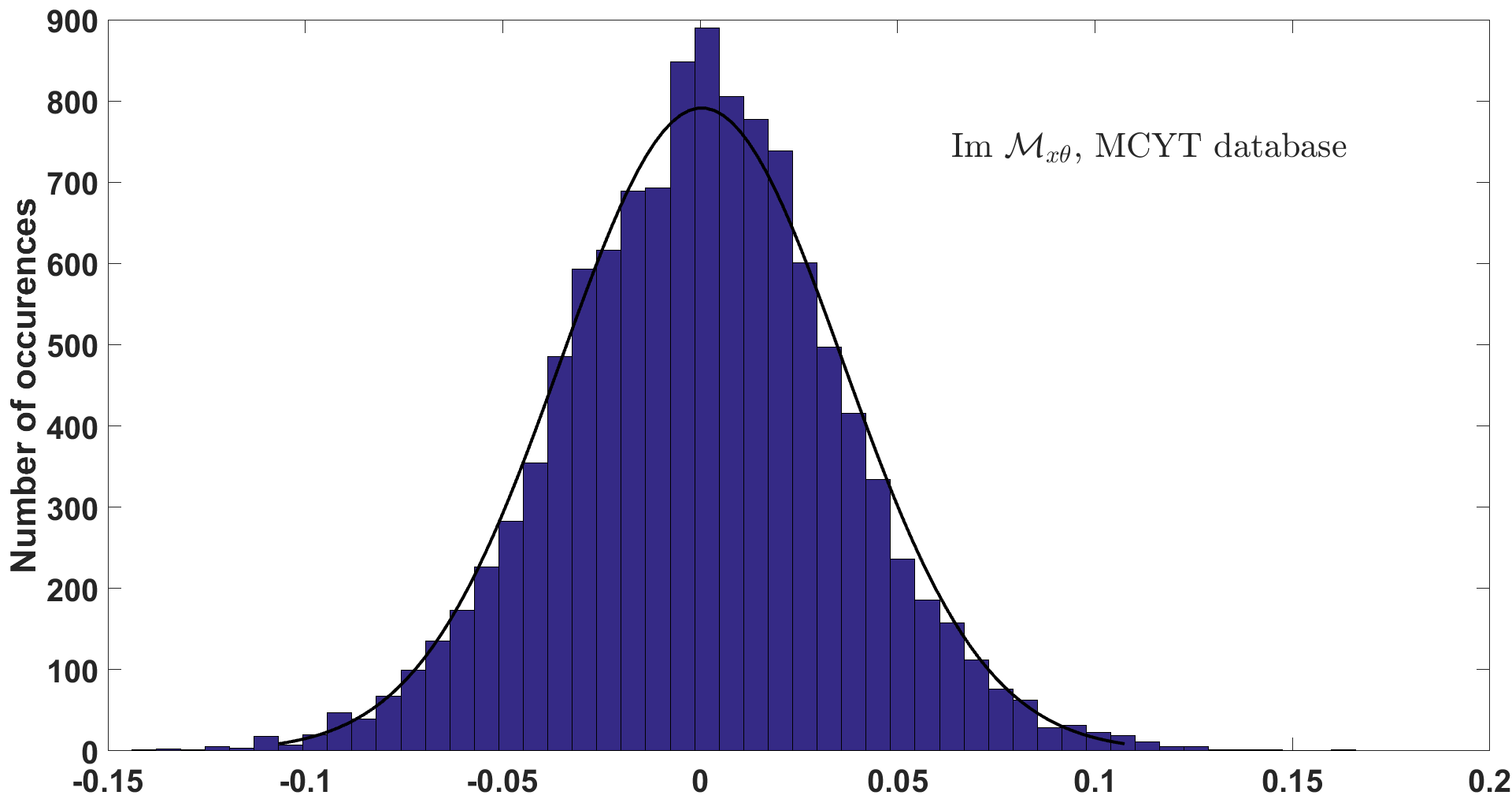}}
\caption{\it Histogram of ${\rm Im}\,\mathcal{M}_{x\theta}$, and a fitted Gaussian.}
	\label{fig:HistogramValueOfMqthetaImaginaryMCYT}
\end{figure}

When using the ZLHDS formulas we will assume that the spectral functions
are Gaussian-distributed.
Figs.\,\ref{fig:HistogramValueOfMqthetaRealMCYT} and \ref{fig:HistogramValueOfMqthetaImaginaryMCYT}
illustrate that this assumption is not far away from the truth.\footnote{
Note that we often see correlations between the real and imaginary part.
This has no influence on the ZLHDS.
}

\subsection{Zero leakage quantization}

\subsubsection{Signal to noise ratio; setting $N$}
\label{sec:SNR}

\begin{figure}[h]
\includegraphics[width=0.49\textwidth]{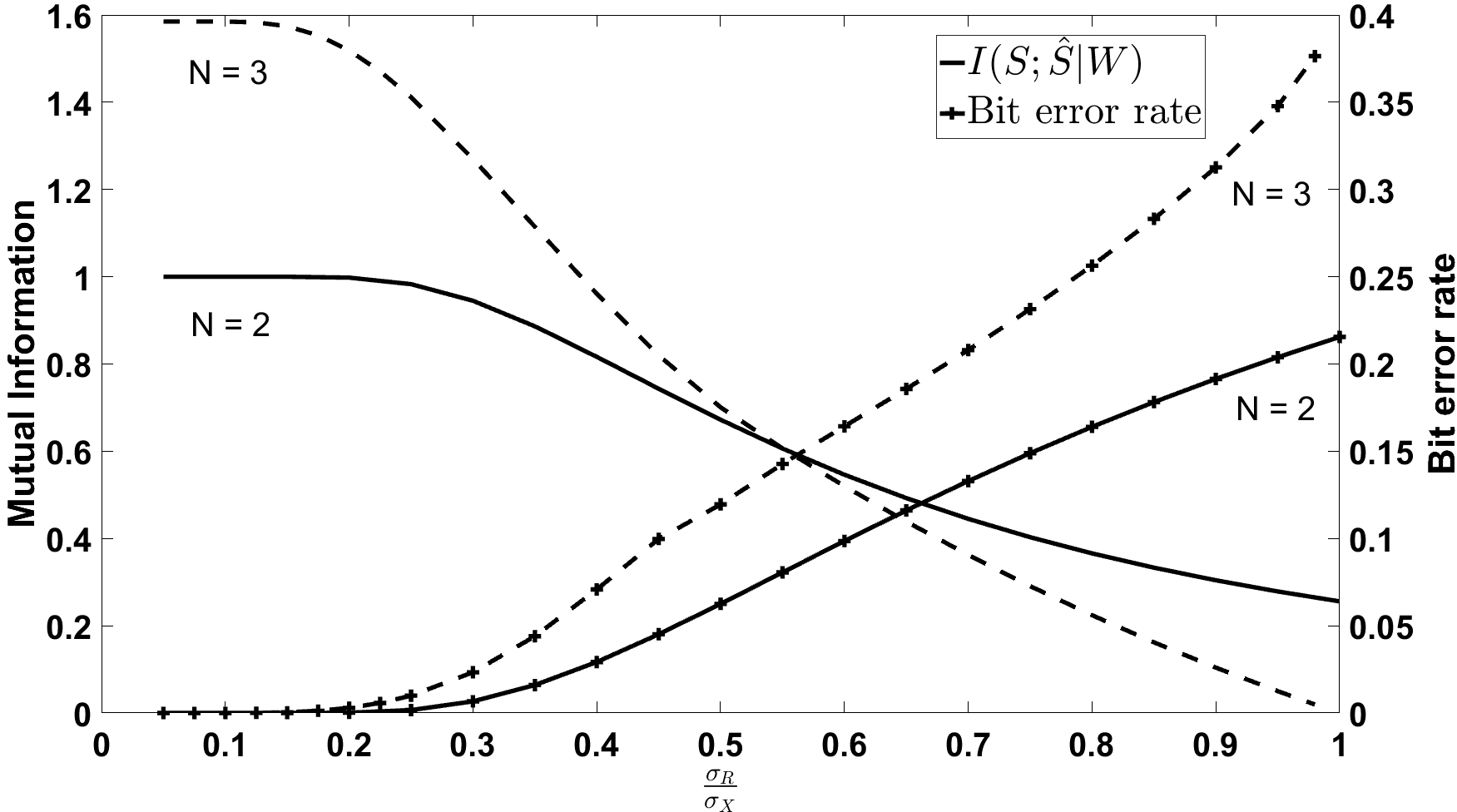}
\caption{\it 
Comparison of ZLHDS with $N=2$ versus $N=3$. 
Lines without markers: Mutual information between the enrolled key $S$ and the reconstructed key $\hat{S}$ given helper data $W$,
as a function of $\sigma_V/\sigma_X$. 
Markers: bit error rate as a function of $\sigma_V/\sigma_X$.
The curves follow
equations (22) and (26) from \cite{SS2017}.
} 
\label{fig:N2vsN3}
\end{figure}

In the ZL HDS of Section~\ref{sec:ZLHDS},
the optimal choice of the parameter $N$ (number of quantization intervals)
depends on the signal to noise ratio.
Fig.\,\ref{fig:N2vsN3} shows a comparison between $N=2$ and $N=3$.
At low noise it is obvious that $N=3$ extracts more information from the source than~$N=2$.
At $\sigma_V/\sigma_X$ larger than approximately $0.3$, there is a regime where
$N=3$ can extract more in theory, but is hindered in practice by the high bit error rate.
At $\sigma_V/\sigma_X>0.55$ the $N=2$ `wins' in all respects.

For our data set, we define a $\sigma_X^2(q,R)$ for every grid point $(q,R)$
as the variance of ${\cal M}(q,R)$ over all images in the database.
The noise $\sigma_V^2(q,R)$ is the variance over all available images of
the same finger, averaged over all fingers.

\begin{figure}
	\centerline{\includegraphics[width=90mm]{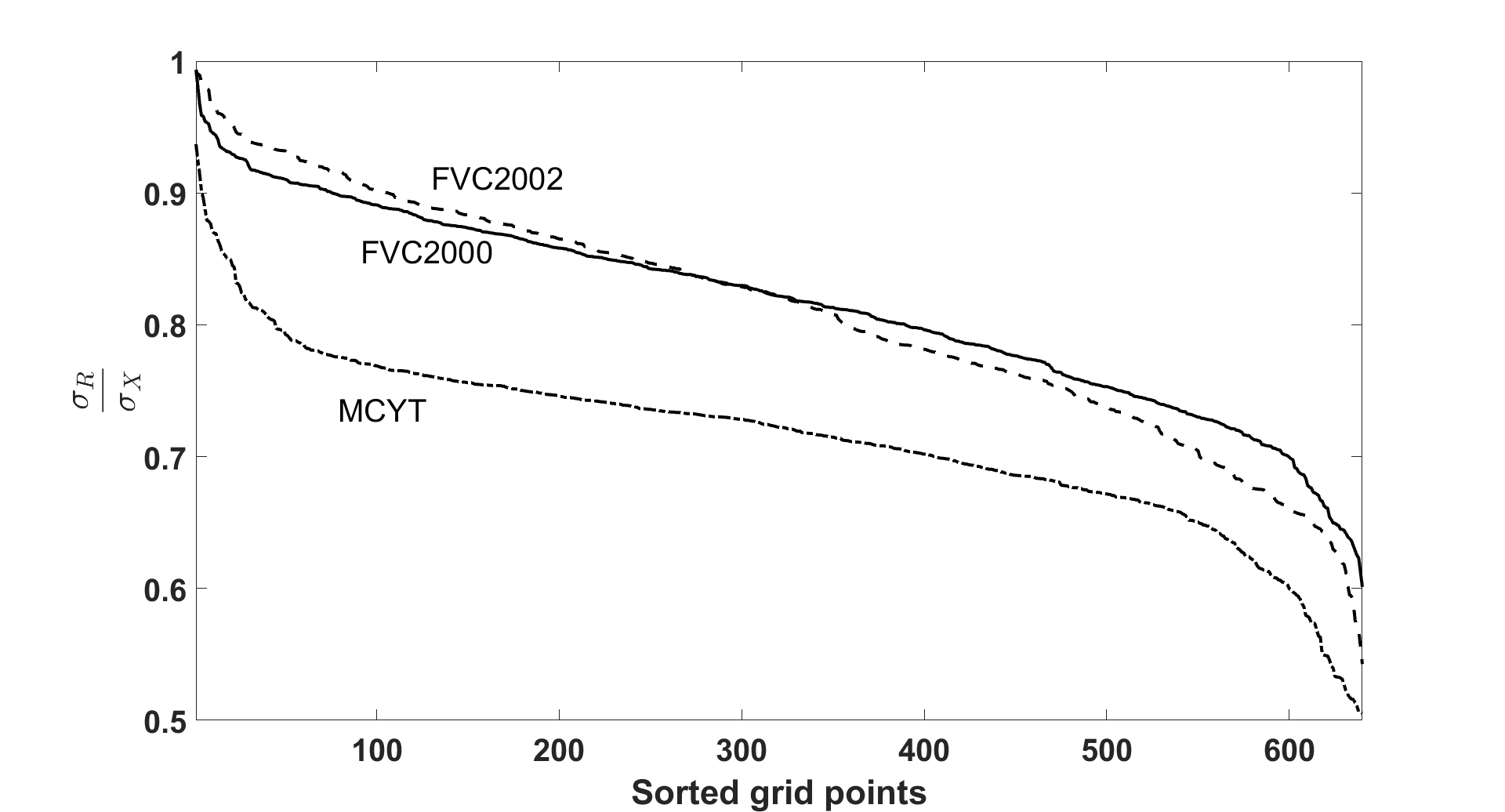}}
	\caption{\it Sorted noise-to-signal  ratio of $\mathcal{M}_{x\theta}$ for different databases.}
	\label{fig:SNR_Theta}
\end{figure}
\begin{figure}
	\centerline{\includegraphics[width=90mm]{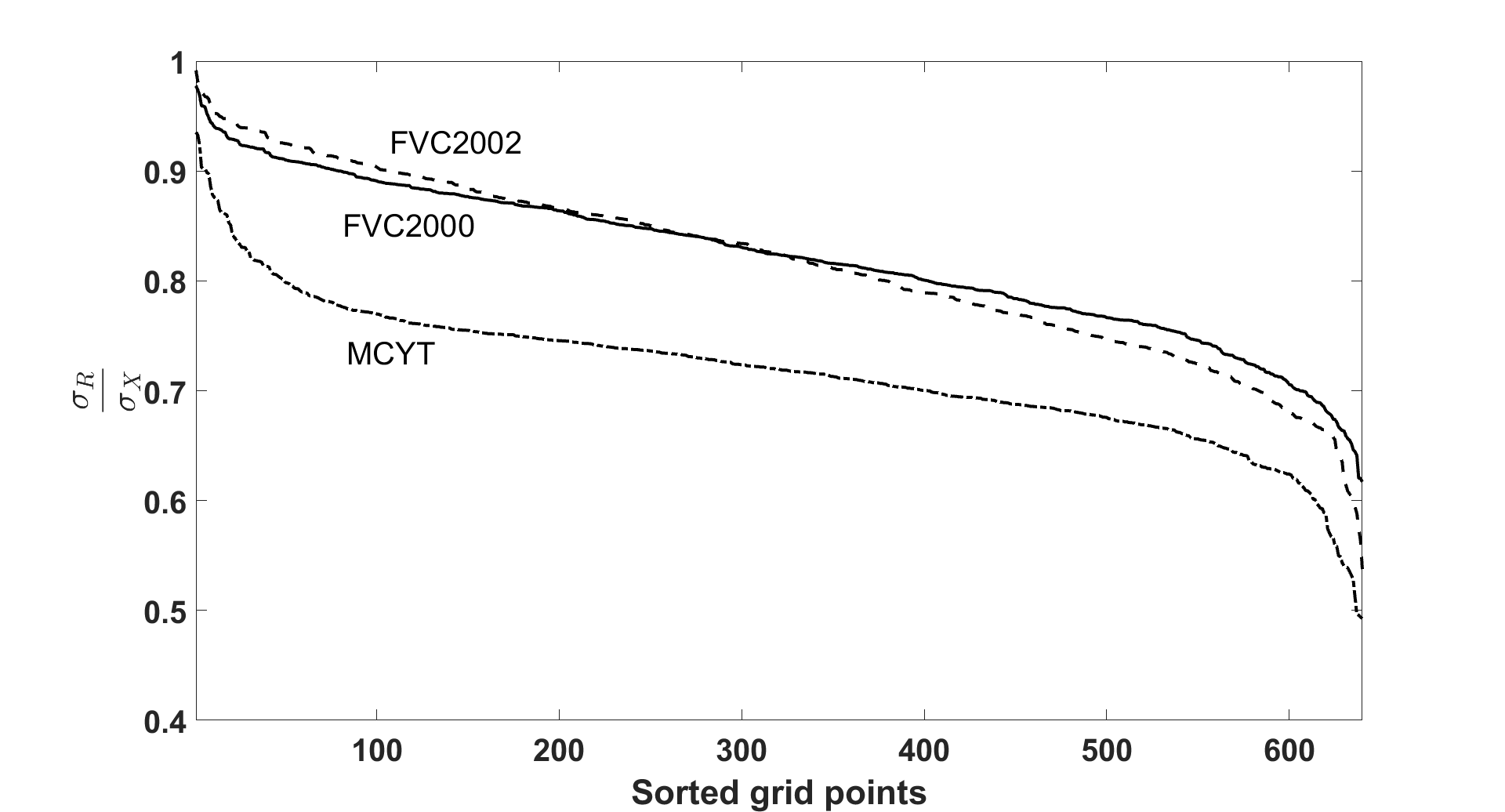}}
		\caption{\it Sorted noise-to-signal  ratio of $\mathcal{M}_{x\beta}$ for different databases.}
	\label{fig:SNR_Beta}
\end{figure}

Figs.\,\ref{fig:SNR_Theta} and \ref{fig:SNR_Beta} show the noise-to-signal ratio. 
Note the large amount of noise; even the best grid points have $\sigma_V/\sigma_X>0.45$.
Fig.\,\ref{fig:N2vsN3} tells us that setting $N=2$ is the best option, and this is the choice we make.
At $N=2$ we extract two bits per grid point from each spectral function (one from ${\rm Re}\,{\mathcal M}$,
one from ${\rm Im}\,{\mathcal M}$). 
Hence our bit string string $k$ (see Fig.\,\ref{fig:twostageHDS}) derived from ${\mathcal M}_{x\theta}$
has length 640.
When we apply fusion of ${\mathcal M}_{x\theta}$ and ${\mathcal M}_{x\beta}$
this becomes 1280.

For $N=2$ the formulas in Section~\ref{sec:ZLHDS} simplify to
${\cal A}_0=(-\infty,0)$, ${\cal A}_1=[0,\infty)$,
$p_0=p_1=\frac12$,
$\xi_{0,w}=F^{\rm inv}(\frac w2)$, $\xi_{1,w}=F^{\rm inv}(\frac12+\frac w2)$,
$\tau_{1,w}=\frac\lambda2(\xi_{0,w}+\xi_{1,w})$.
Since we work with Gaussian distributions, $F$ is the Gaussian cdf (`probability function').
\subsubsection{Enrollment and reconstruction}
\label{sec:enroll}

We have experimented with three different enrollment methods:\\
E1. A single image is used. \\
E2: We take the first\footnote{
We take the {\em first} $t$ images to show that the approach works. We are not trying to optimise the choice of images.
} 
$t$ images of a finger and calculate the average spectral function.
We call this the `superfinger' method.
In the ZLHDS calculations the signal-to-noise ratio of the average spectral function is used.

E3: For each of $t$ images  we calculate an enrollment string~$k$. 
We apply bitwise majority voting on these strings. (This requires odd $t$.)
The reconstruction boundaries are calculated based on the superfinger method, i.e. as in E2.

{\it Reconstruction:}\\
We study fingerprint authentication with genuine pairs and impostor pairs. 
For pedagogical reasons we will present results at each stage of the signal processing:
(1) spectral function domain, before quantisation;
(2) binarized domain, without HDS;
(3) with ZLHDS;
(4) with ZLHDS and discarding the highest-noise grid points.

In the spectral function domain the fingerprint matching is done via a correlation score \cite{SS2017}.
In the binarized domain we look at the Hamming weight between the enrolled $k$ and the reconstructed~$\hat k$.
For all cases we will show ROC curves in order to visualise the FAR-FRR tradeoff as a function
of the decision threshold.

Let the number of images per finger 
be denoted as $M$, and the number of fingers in a database as~$L$ .\\

E1: For the spectral domain and the quantization without HDS
we compare all genuine pairs, i.e. ${M\choose 2}$  image pairs per finger, resulting in $L{M \choose 2}$ data points. 
For ZLHDS the number is twice as large, since there is an asymmetry between enrollment and reconstruction.
For the FVC databases we generate all possible impostor combinations (all images of all impostor fingers),
resulting in ${\cal O}(M^2 L^2)$ data points.

For the MCYT database, which is larger,
we take only {\em one random image} per impostor finger, resulting in ${\cal O}(ML^2)$ data points.\\
E2+E3:  For genuine pairs we compare the superfinger to the remaining $M-t$ images. 
Thus we have $(M-t)L$ data points. 
Impostor pairs are generated as for E1.

Note: The VeriFinger software
was not able to extract information for every image. 
\section{Experimental results}
\label{sec:results}

\subsection{FAR/FRR rates before error correction}

\begin{figure}[t]
	\centerline{\includegraphics[width=90mm]{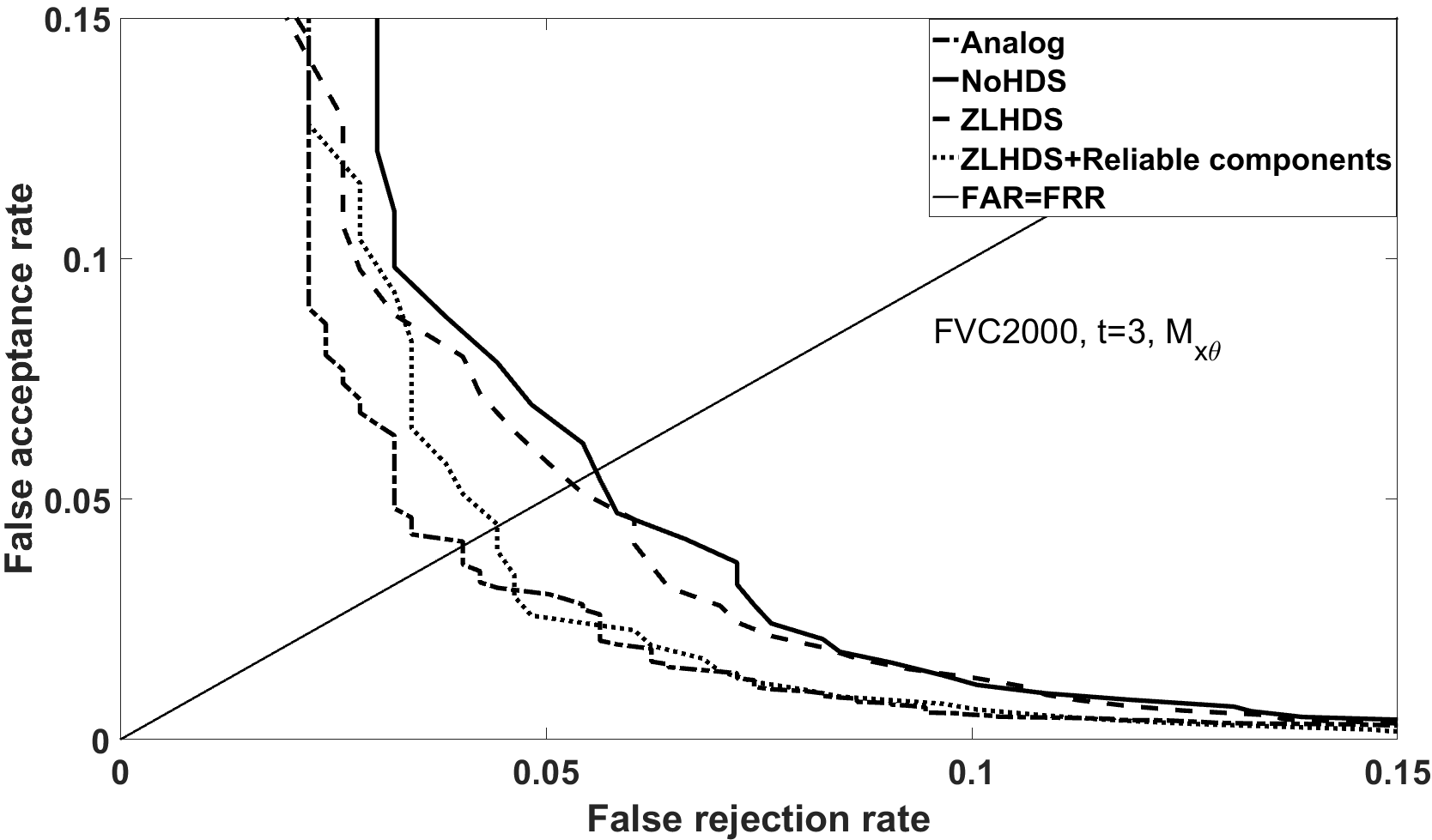}}
	\centerline{\includegraphics[width=90mm]{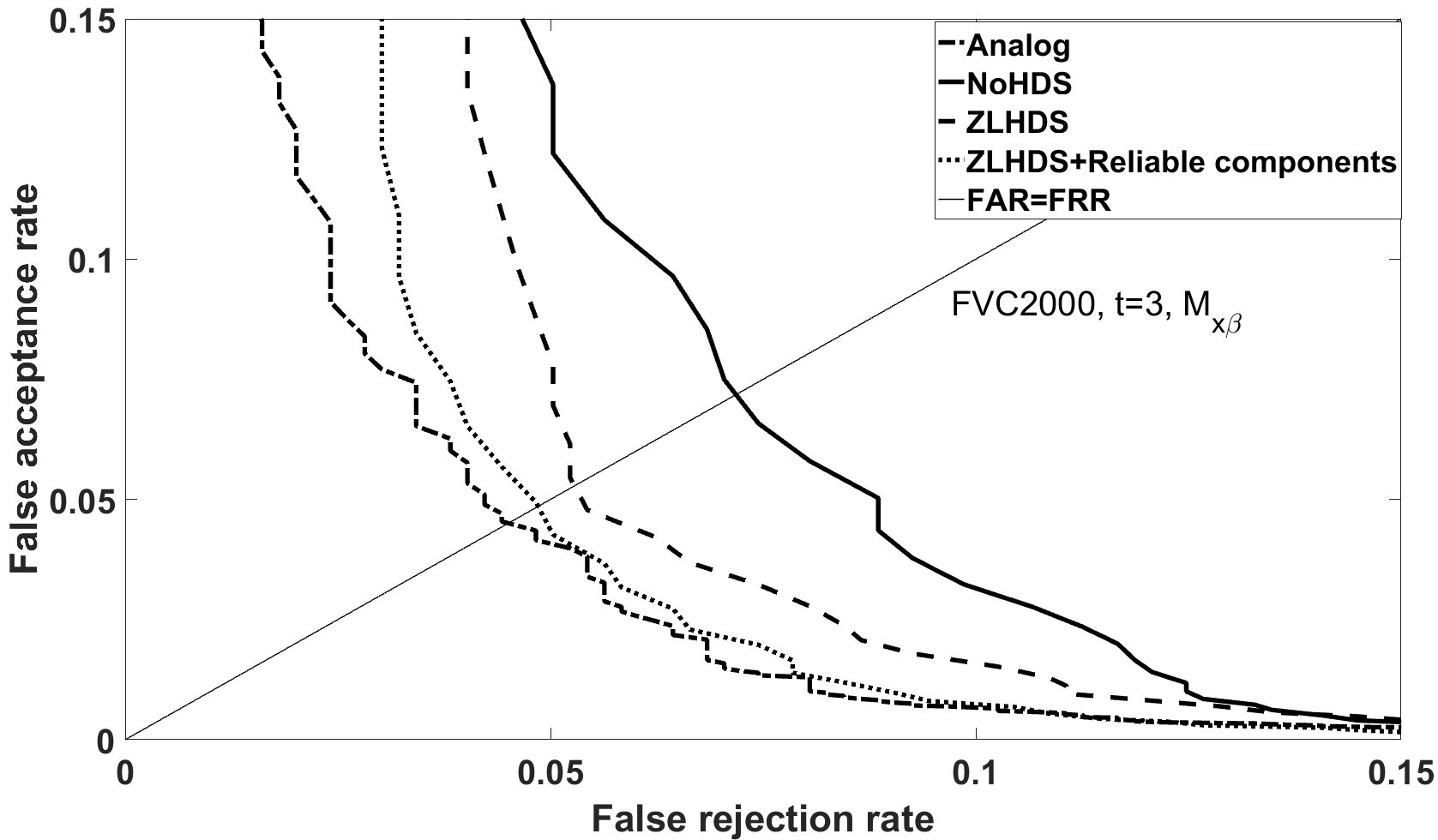}}
	\centerline{\includegraphics[width=90mm]{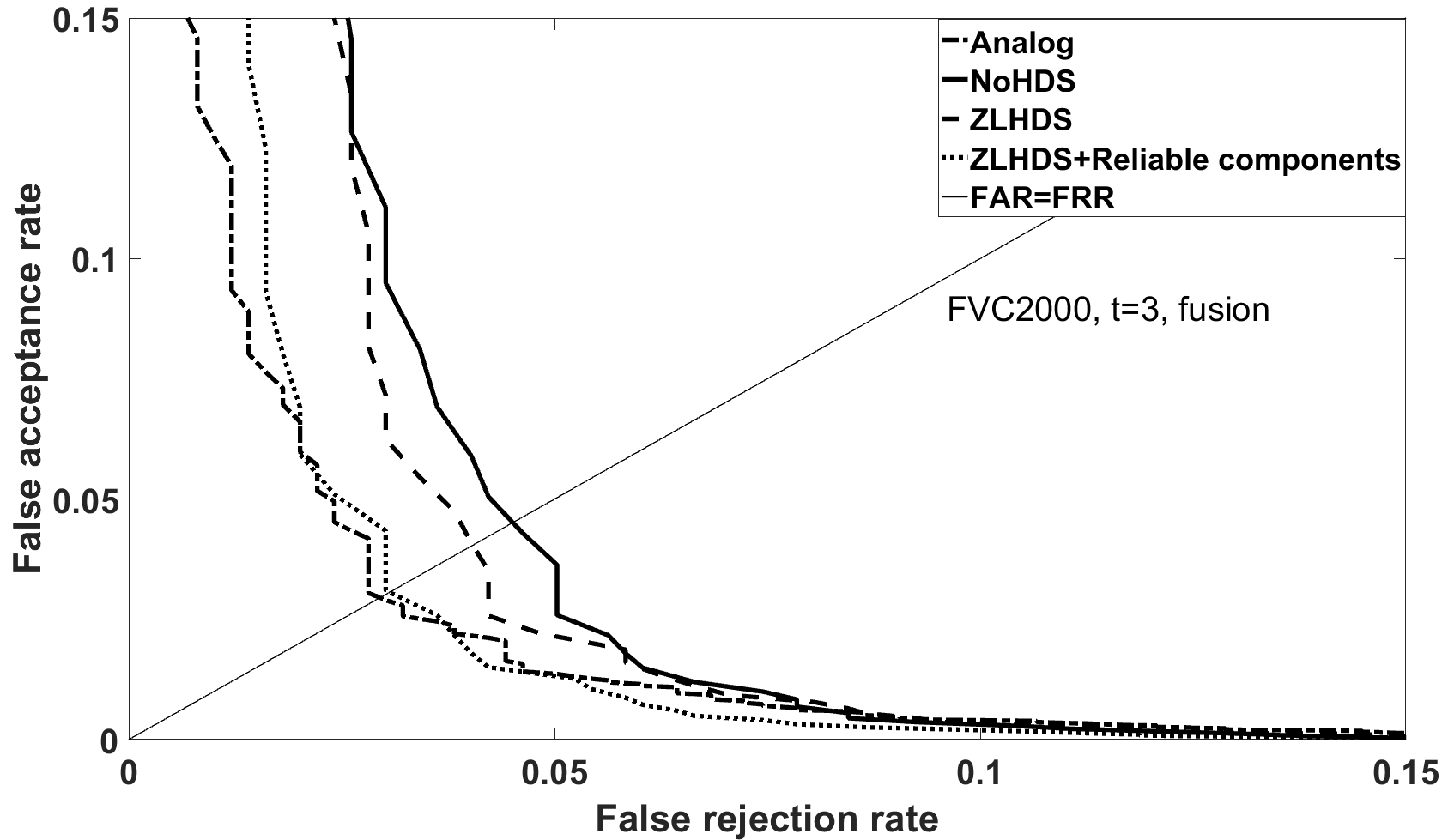}}
\caption{\it Performance result for several processing methods. FVC2000. Enrollment method E2 with $t=3$.}
\label{fig:ROCstagesFVC}
\end{figure}

\begin{figure}[h]
\centerline{\includegraphics[width=90mm]{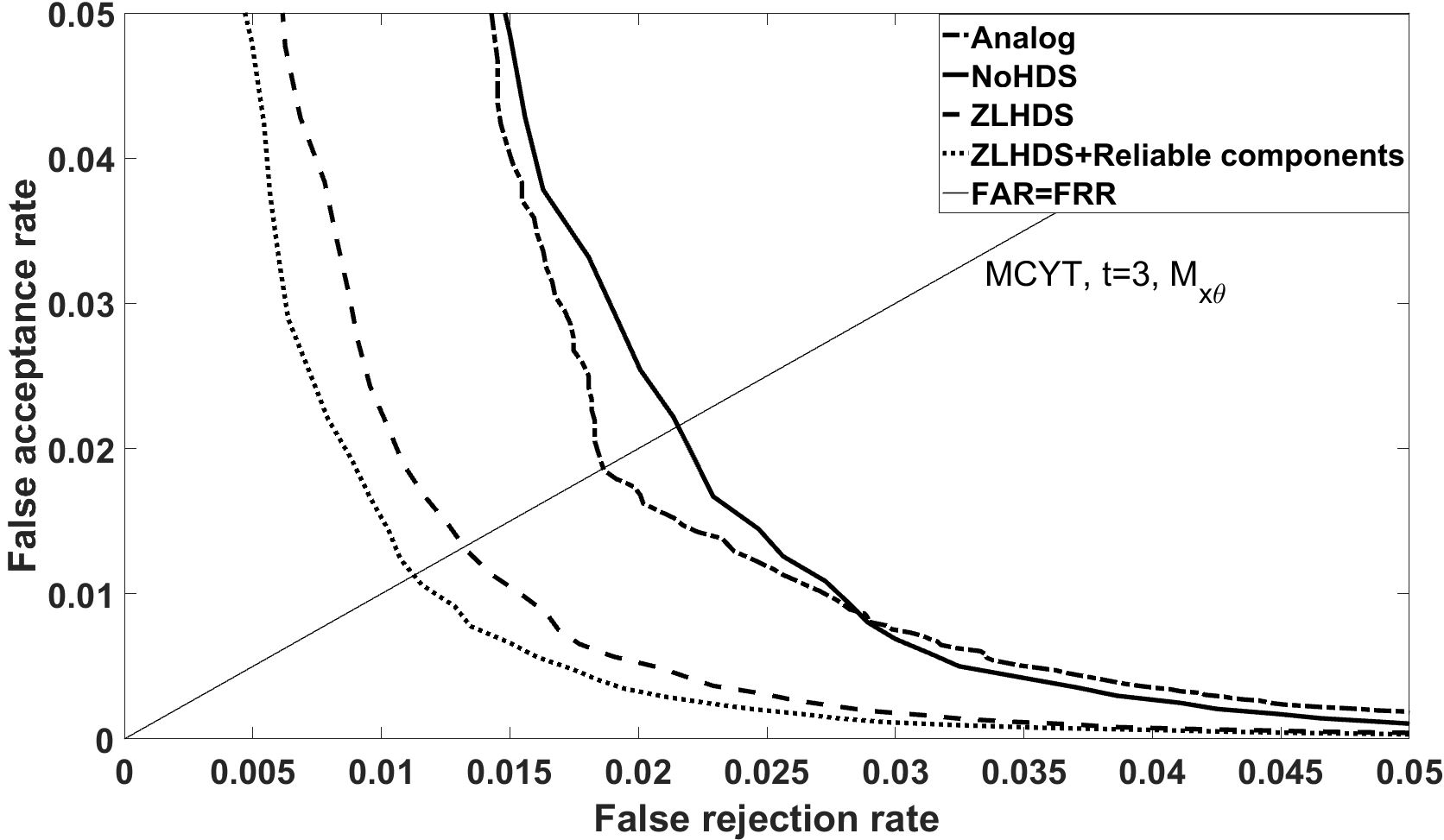}}
\centerline{\includegraphics[width=90mm]{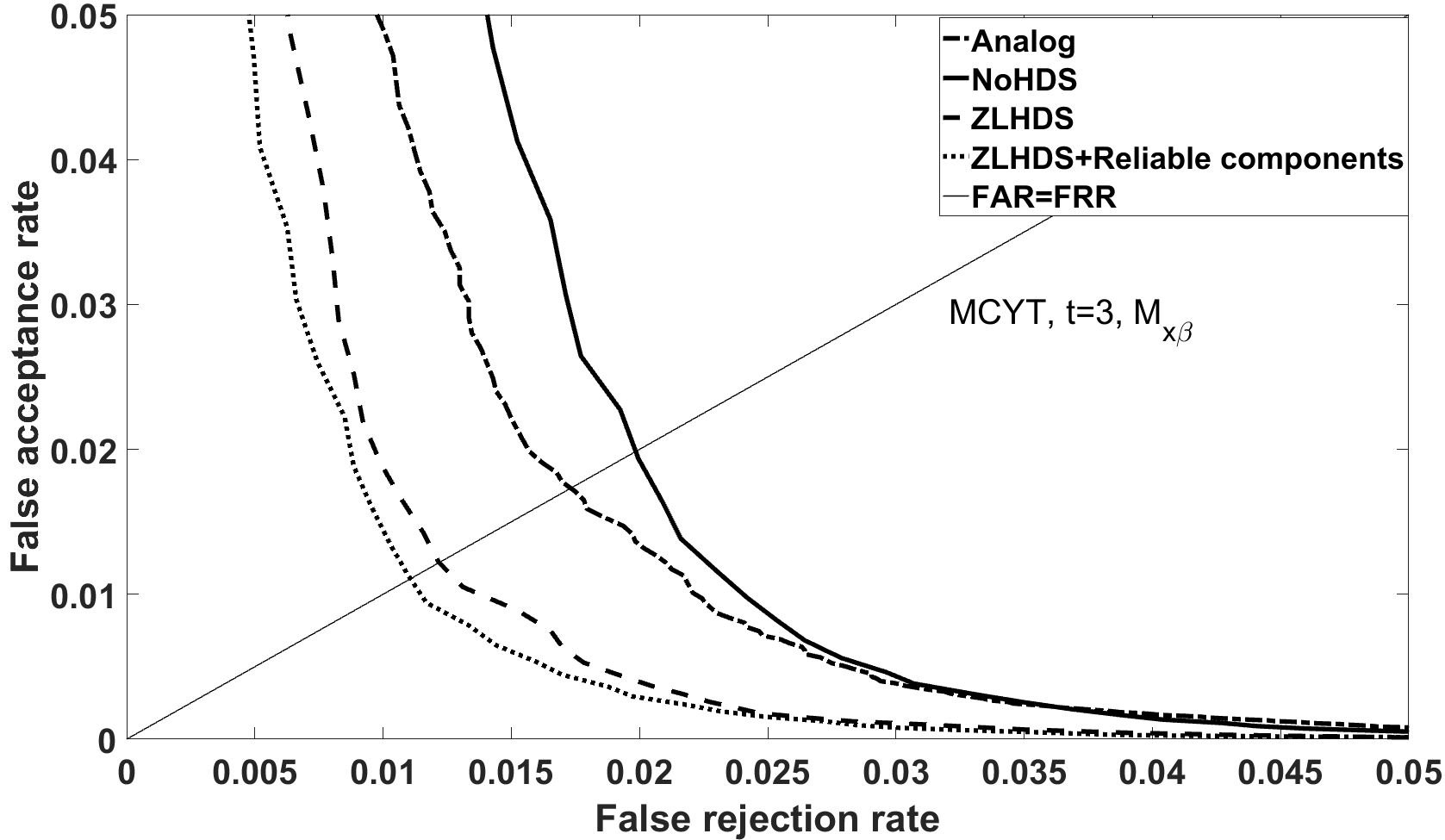}}
\centerline{\includegraphics[width=90mm]{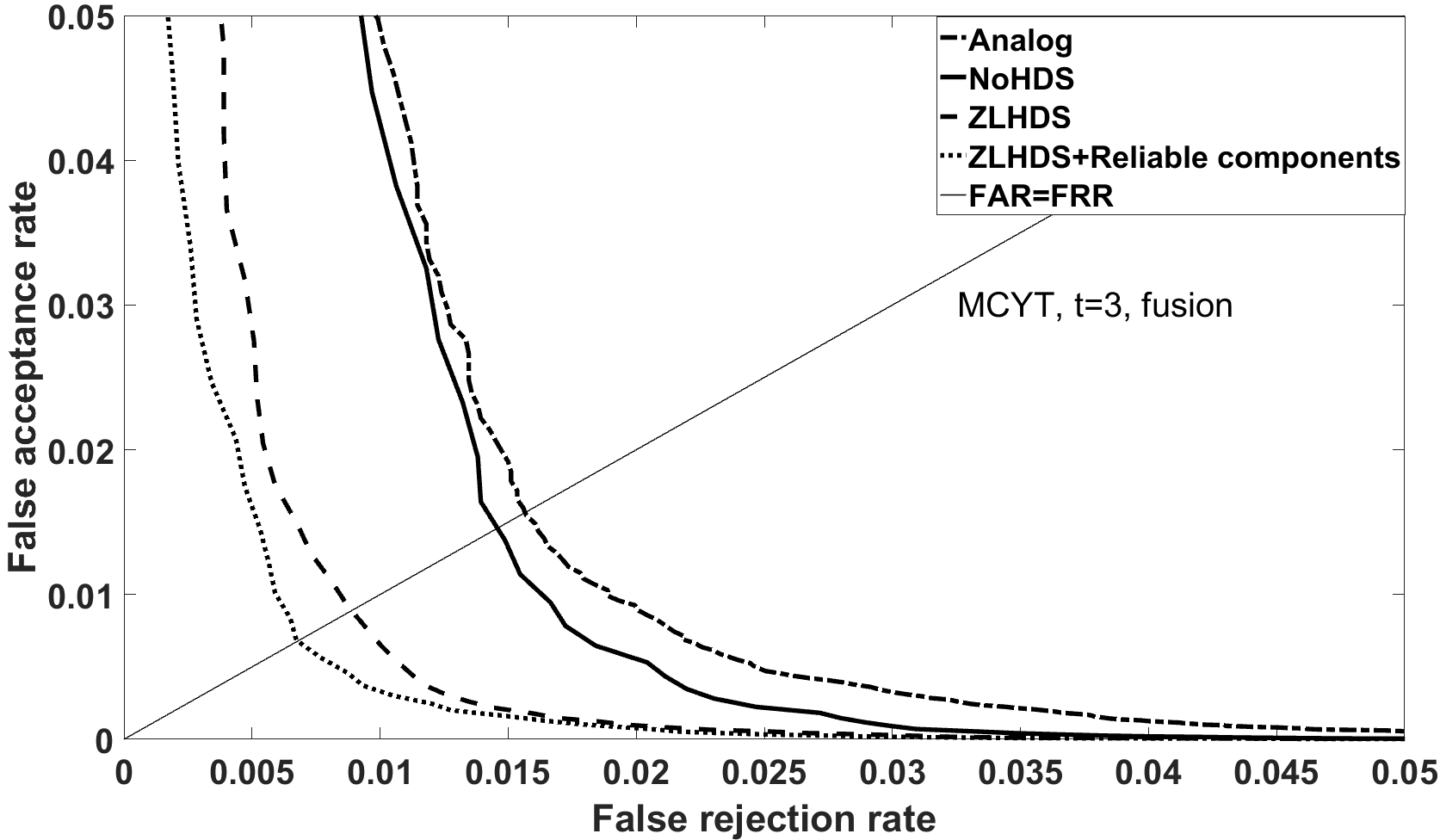}}
\caption{\it Performance result for several processing methods. MCYT. Enrollment method E2 with $t=3$.}
\label{fig:ROCstagesMCYT}
\end{figure}

\begin{figure}[h]
\centerline{\includegraphics[width=90mm]{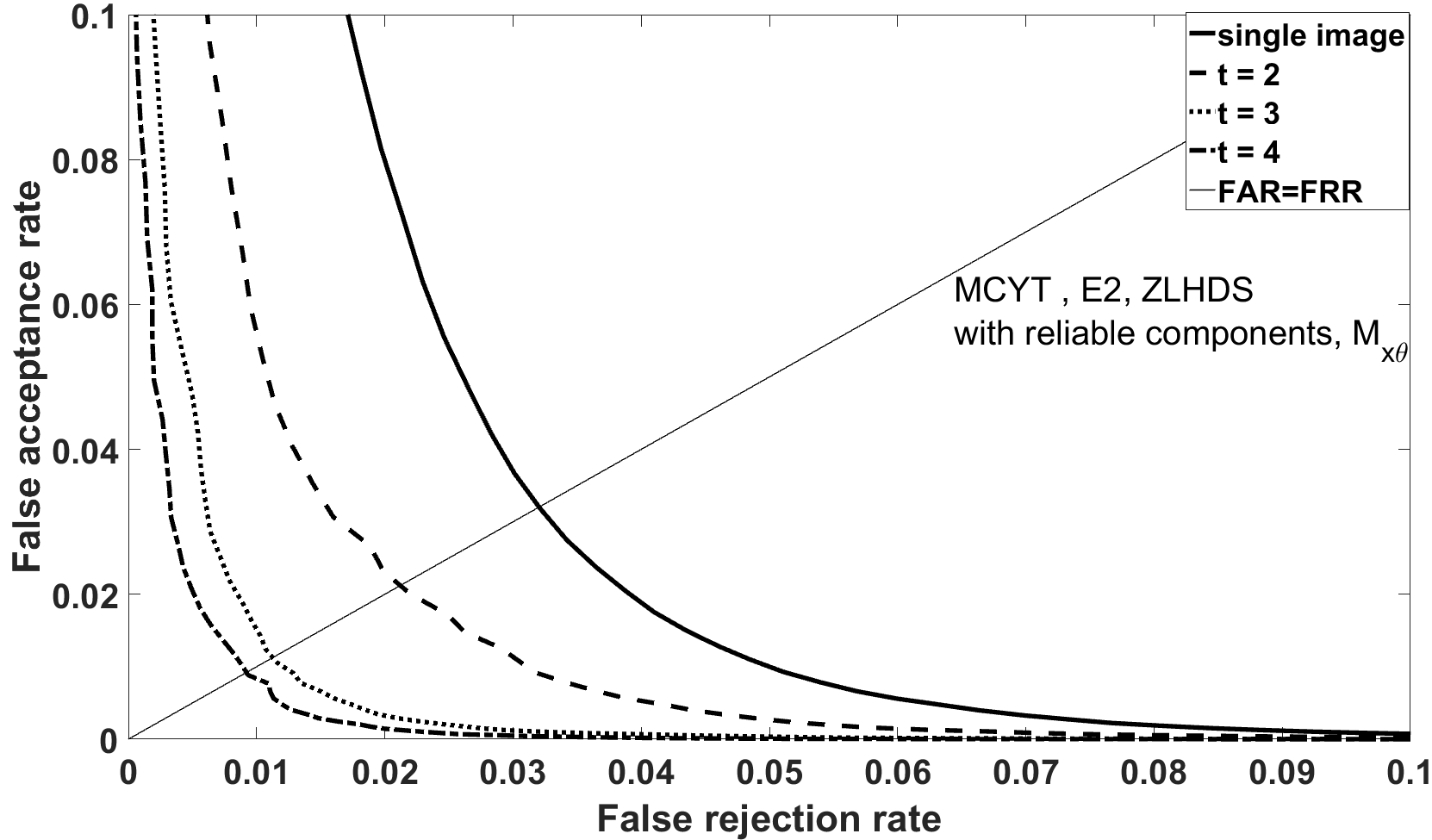}}
\centerline{\includegraphics[width=90mm]{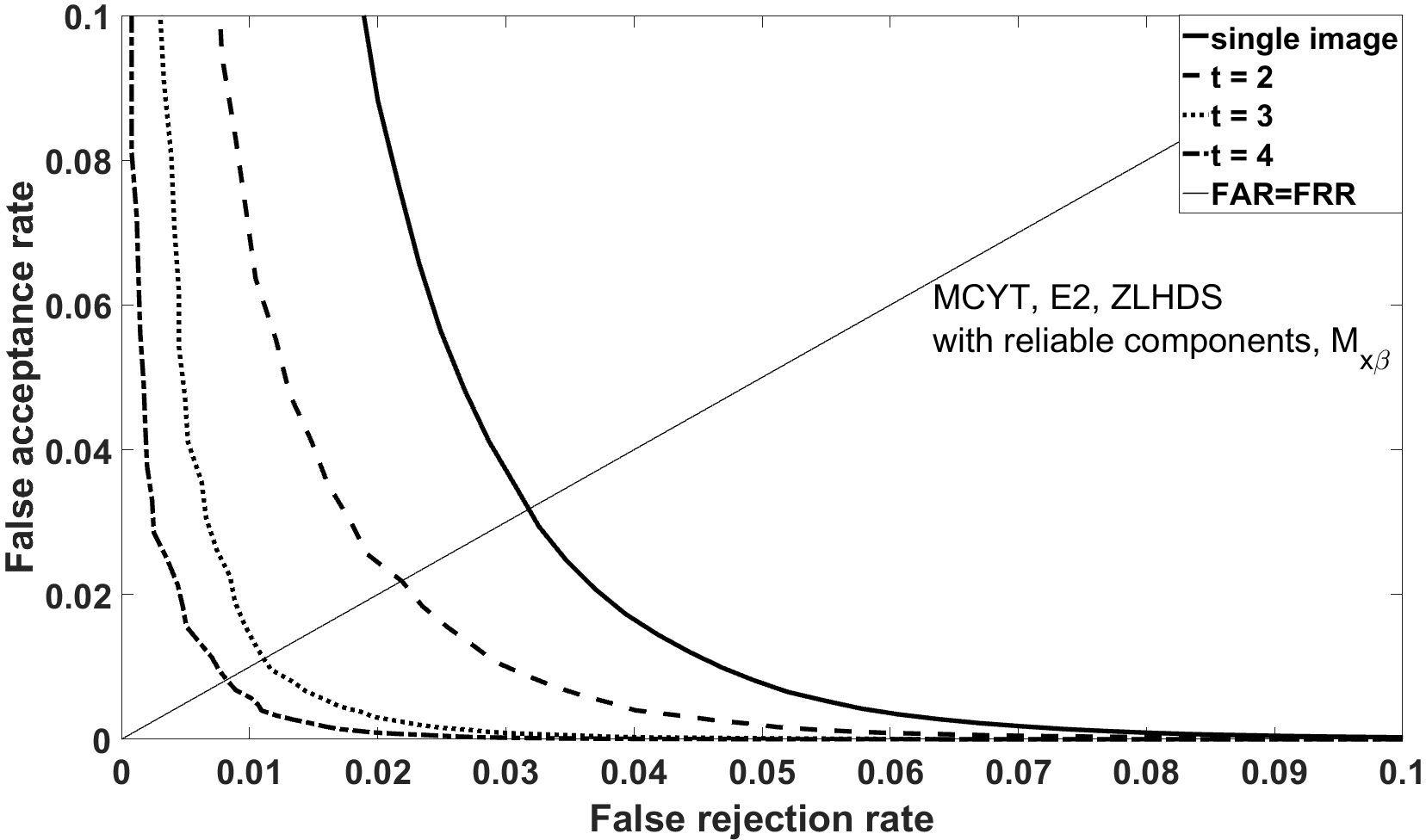}}
\centerline{\includegraphics[width=90mm]{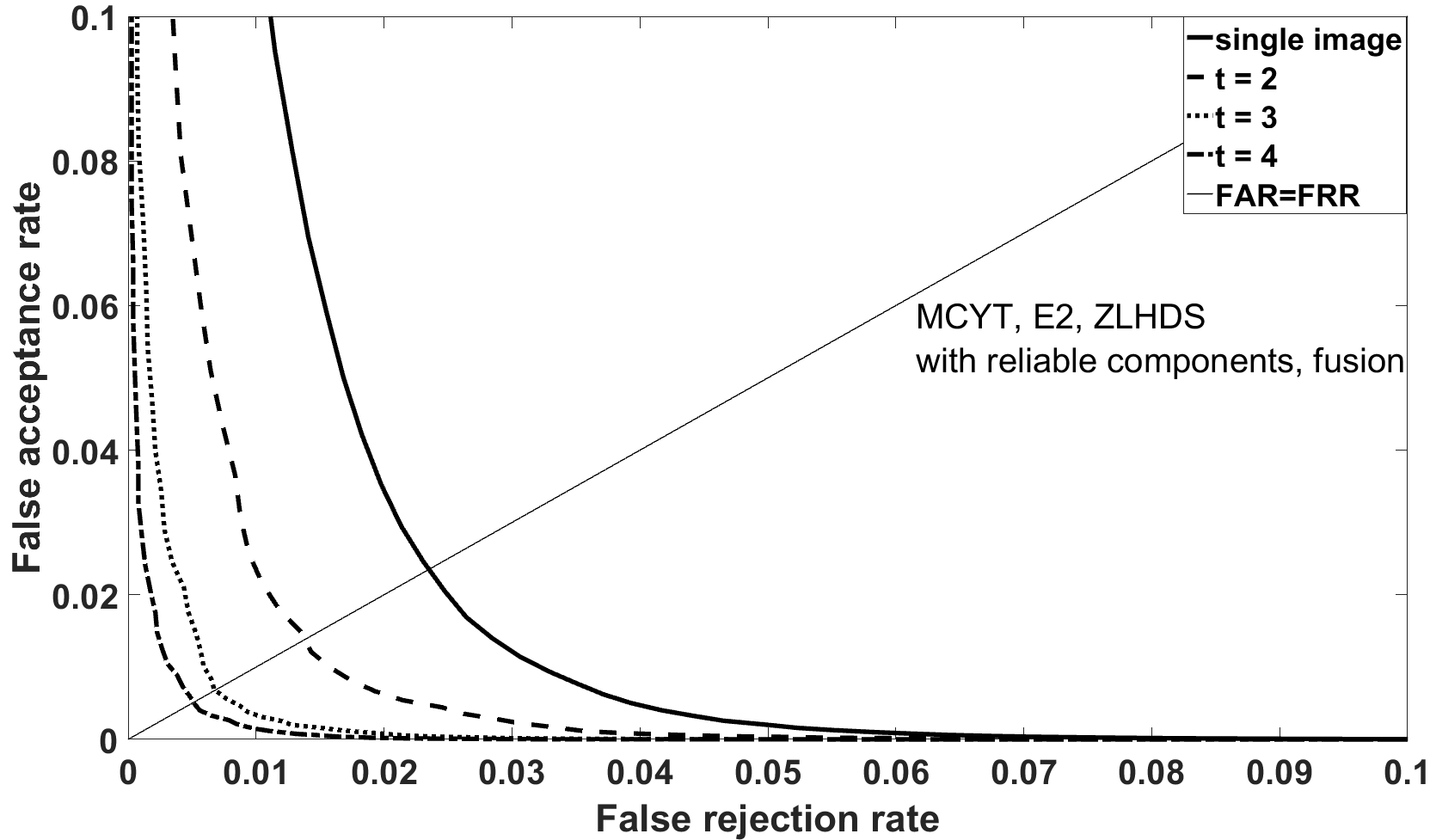}}
\caption{\it Performance effect of the number of images used for enrollment.}
\label{fig:ROCt}
\end{figure}

\begin{figure}[h]
\centerline{\includegraphics[width=90mm]{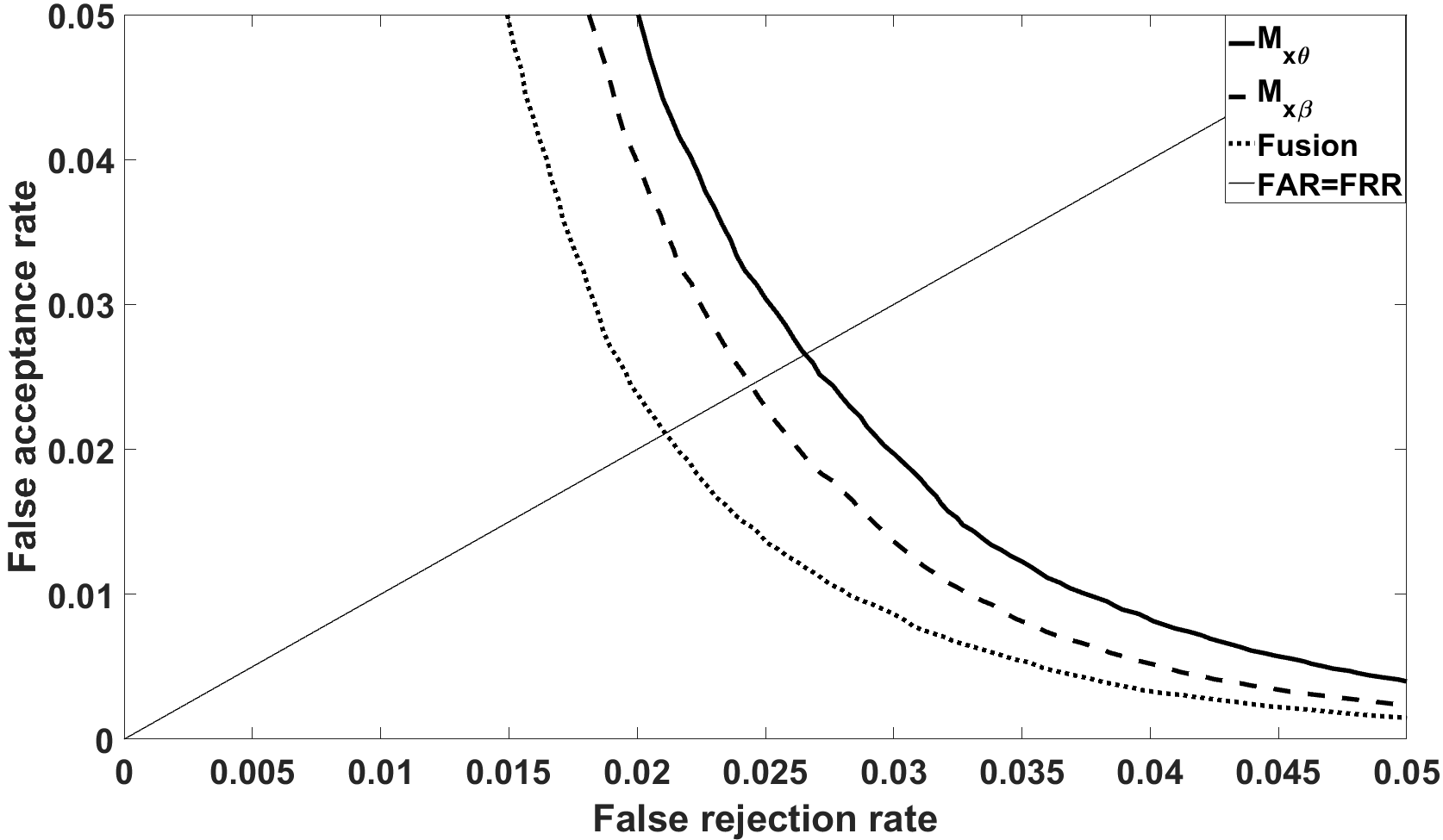}}
\caption{\it Performance of $M_{x\theta}$ and $M_{x\beta}$ individually, and of their fusion.
 	MCYT database; enrollment method E1; analog domain.}
\label{fig:fusion}
\end{figure}

For each the data processing steps/options before application
of the Code Offset method,
we investigate the False Accept rates and False Reject rates.
We identify a number of trends.
\begin{itemize}[leftmargin=4mm,itemsep=0mm]
\item
Figs.\,\ref{fig:ROCstagesFVC} and \ref{fig:ROCstagesMCYT} show ROC curves.
All the non-analog curves were made under the implicit assumption 
that for each decision threshold (number of bit flips)
an error-correcting code
can be constructed that enforces that threshold, i.e. decoding succeeds only if the number
of bit flips is below the threshold.
Unsurprisingly, we see in the figures that quantisation causes a performance penalty.
Furthermore the penalty is clearly less severe when the ZLHDS is used.
Finally, it is advantageous to discard some grid points that have bad signal-to-noise ratio.
For the curves labeled `ZLHDS+reliable components' only the least noisy\footnote{
This is defined as a global property of the whole database.
The selection of reliable components does {\em not} reveal anything about an individual.
Note that \cite{Nanda2010} does reveal personalised reliable components
and obtains better FA and FN error rates.
}
512 bits of $k$ were kept
(1024 in the case of fusion). Our choice for the number 512 is not entirely arbitrary: it fits
error-correcting codes.
Note in Fig.\,\ref{fig:ROCstagesMCYT} that ZLHDS with reliable component selection
performs better than analog spectral functions {\em without} reliable component selection.
(But not better than analog with selection.)
\item
The E2 and E3 enrollment methods perform better than E1.
Furthermore, performance increases with~$t$.
A typical example is shown in Fig.\,\ref{fig:ROCt}.
\item
The spectral functions ${\cal M}_{x\theta}$ and ${\cal M}_{x\beta}$
individually have roughly the same performance.
Fusion yields a noticeable improvement.
An example is shown in Fig.\,\ref{fig:fusion}. 
(We implemented fusion in the analog domain as addition of the two similarity scores.)
\item
Tables \ref{TableMCYT} to \ref{TableMCYTMV} show Equal Error Rates and Bit Error Rates.
We see that enrollment methods E2 and E3 have similar performance,
with E2 yielding a somewhat lower genuine-pair BER than E3.
\item
In Table~\ref{TableMCYT} it may look strange that the EER
in the rightmost column is sometimes lower than in the `analog' column. 
We think this happens because there is no reliable component selection in 
the `analog' procedure.
\item
Ideally the impostor BER is 50\%.
In the tables we see that the impostor BER can get lower than 50\% when the ZLHDS is used
and the enrollment method is~E2. On the other hand, it is always around 50\% 
in the `No HDS' case.
This seems to contradict the Zero Leakage property of the helper data system.
The ZLHDS is supposed not to leak, i.e. the helper data should not help impostors.
However,
the zero-leakage property is guaranteed to hold only if the variables are independent.
In real-life data there are correlations between grid points and correlations
between the real and imaginary part of a spectral function.

\end{itemize}
 

\begin{table}[h]
	\caption{\it Equal Error Rates and Bit Error Rates. MCYT database. Enrollment methods E1 and E2. 
	Numbers displayed as a percentage are EERs. 
	Numbers without a \% sign are BERs: the left number is for genuine pairs, right for impostors.}
	\begin{center}
		\begin{tabular}{|c|c|c|c|c c|c c|c c|} 
			\hline
			\multicolumn{2}{|c|}{\multirow{2}{*}{\#images ($t$)}}&	\multicolumn{2}{|c|}{\multirow{2}{*}{Analog}}&	\multicolumn{2}{|c|}{\multirow{2}{*}{No HDS}}&	\multicolumn{2}{|c|}{\multirow{2}{*}{ZLHDS}}&	\multicolumn{2}{|c|}{\multirow{2}{*}{ZLHDS+r.c.}}\\
			\multicolumn{2}{|c|}{}&	\multicolumn{2}{|c|}{\multirow{2}{*}{}}&	\multicolumn{2}{|c|}{\multirow{2}{*}{}}&	\multicolumn{2}{|c|}{\multirow{2}{*}{}}&	\multicolumn{2}{|c|}{\multirow{2}{*}{}}\\
			\hline 
			\multicolumn{1}{|c|}{\multirow{6}{*}{1}}&\multicolumn{1}{|c|}{\multirow{2}{*}{${\cal M}_{x\theta}$}}&	\multicolumn{2}{|c|}{\multirow{2}{*}{2.6\% }}&	\multicolumn{2}{|c|}{3.7\%}&	\multicolumn{2}{|c|}{3.4\%}&	\multicolumn{2}{|c|}{3.2\%}\\
			\multicolumn{1}{|c|}{}&	\multicolumn{1}{|c|}{}&	\multicolumn{2}{|c|}{\multirow{2}{*}{}}&0.33 & 0.50&	0.30	& 0.49&	0.29 & 0.49			\\
			\cline{2-10}
			\multicolumn{1}{|c|}{}&\multicolumn{1}{|c|}{\multirow{2}{*}{${\cal M}_{x\beta}$}}&	\multicolumn{2}{|c|}{\multirow{2}{*}{2.4\%}}&	\multicolumn{2}{|c|}{3.7\%}&	\multicolumn{2}{|c|}{3.4\%}&	\multicolumn{2}{|c|}{3.2\%}\\
			\multicolumn{1}{|c|}{}&	\multicolumn{1}{|c|}{}&	\multicolumn{2}{|c|}{\multirow{2}{*}{}}&	0.33 & 0.50 &	0.31	& 0.50 &	0.29 & 0.49		\\
			\cline{2-10}
			\multicolumn{1}{|c|}{}&\multicolumn{1}{|c|}{\multirow{2}{*}{Fusion}}&	\multicolumn{2}{|c|}{\multirow{2}{*}{2.1\%}}&	\multicolumn{2}{|c|}{2.9\%}&	\multicolumn{2}{|c|}{2.6\%}&	\multicolumn{2}{|c|}{2.3\%}\\
			\multicolumn{1}{|c|}{}&	\multicolumn{1}{|c|}{}&		\multicolumn{2}{|c|}{\multirow{2}{*}{}}&	0.33 & 0.50&0.30	& 0.49 		&	0.29 	& 0.49 \\
			\hline 
			\multicolumn{1}{|c|}{\multirow{6}{*}{2}}&\multicolumn{1}{|c|}{\multirow{2}{*}{${\cal M}_{x\theta}$}}&		\multicolumn{2}{|c|}{\multirow{2}{*}{2.1\%}}&	\multicolumn{2}{|c|}{3.2\%}&	\multicolumn{2}{|c|}{2.3\%	}&	\multicolumn{2}{|c|}{2.1\%}\\
			\multicolumn{1}{|c|}{}&	\multicolumn{1}{|c|}{}&		\multicolumn{2}{|c|}{\multirow{2}{*}{}}&	0.33	& 0.50 & 0.28	& 0.46 	&	0.27 	& 0.46\\
			\cline{2-10}
			\multicolumn{1}{|c|}{}&\multicolumn{1}{|c|}{\multirow{2}{*}{${\cal M}_{x\beta}$}}&	\multicolumn{2}{|c|}{\multirow{2}{*}{1.7\%}}&	\multicolumn{2}{|c|}{3.01\%}&	\multicolumn{2}{|c|}{2.4\%}&	\multicolumn{2}{|c|}{2.2\%}\\
			\multicolumn{1}{|c|}{}&	\multicolumn{1}{|c|}{}&	\multicolumn{2}{|c|}{\multirow{2}{*}{}}& 0.33 & 0.50	& 0.28 & 0.47 &	0.27 & 0.47	\\
			\cline{2-10}
			\multicolumn{1}{|c|}{}&\multicolumn{1}{|c|}{\multirow{2}{*}{Fusion}}&	\multicolumn{2}{|c|}{\multirow{2}{*}{1.6\%}}&	\multicolumn{2}{|c|}{2.3\%}&	\multicolumn{2}{|c|}{1.7\% }&	\multicolumn{2}{|c|}{1.4\%} \\
			\multicolumn{1}{|c|}{}&	\multicolumn{1}{|c|}{}&	\multicolumn{2}{|c|}{\multirow{2}{*}{}}&	0.33 & 0.50 &	0.28	& 0.46 & 0.27 & 0.47\\
			\hline 
			\multicolumn{1}{|c|}{\multirow{6}{*}{3}}&\multicolumn{1}{|c|}{\multirow{2}{*}{${\cal M}_{x\theta}$}}&	\multicolumn{2}{|c|}{\multirow{2}{*}{1.4\%}}&	\multicolumn{2}{|c|}{2.2\%}&	\multicolumn{2}{|c|}{1.3\%}&	\multicolumn{2}{|c|}{1.1\%}\\
			\multicolumn{1}{|c|}{}&	\multicolumn{1}{|c|}{}&	\multicolumn{2}{|c|}{\multirow{2}{*}{}}&	0.31 	& 0.50	& 0.24 & 0.45 & 0.23 & 0.46 \\
			\cline{2-10}
			\multicolumn{1}{|c|}{}&\multicolumn{1}{|c|}{\multirow{2}{*}{${\cal M}_{x\beta}$}}&	\multicolumn{2}{|c|}{\multirow{2}{*}{1.1\%}}&	\multicolumn{2}{|c|}{2.0\%}&	\multicolumn{2}{|c|}{1.2\%}&	\multicolumn{2}{|c|}{1.1\%}\\
			\multicolumn{1}{|c|}{}&	\multicolumn{1}{|c|}{}&	\multicolumn{2}{|c|}{\multirow{2}{*}{}}&0.31 & 0.50 & 0.25 & 0.46 & 0.23 & 0.46 \\
			\cline{2-10}
			\multicolumn{1}{|c|}{}&\multicolumn{1}{|c|}{\multirow{2}{*}{Fusion}}&		\multicolumn{2}{|c|}{\multirow{2}{*}{1.1\%}}&	\multicolumn{2}{|c|}{1.5\%}&	\multicolumn{2}{|c|}{0.9\%}&	\multicolumn{2}{|c|}{0.7\%}\\
			\multicolumn{1}{|c|}{}&	\multicolumn{1}{|c|}{}&	\multicolumn{2}{|c|}{\multirow{2}{*}{}}& 0.31 & 0.50 &	0.24 & 0.46 &	0.23 & 0.46 \\
			\hline 
			\multicolumn{1}{|c|}{\multirow{6}{*}{4}}&\multicolumn{1}{|c|}{\multirow{2}{*}{${\cal M}_{x\theta}$}}&	\multicolumn{2}{|c|}{\multirow{2}{*}{1.2\%}}&	\multicolumn{2}{|c|}{1.7\%}&	\multicolumn{2}{|c|}{1.0\%}&	\multicolumn{2}{|c|}{0.9\%}\\
			\multicolumn{1}{|c|}{}&	\multicolumn{1}{|c|}{}&	\multicolumn{2}{|c|}{\multirow{2}{*}{}}&	0.29 & 0.50 & 0.22 & 0.45 & 0.21 & 0.45 \\
			\cline{2-10}
			\multicolumn{1}{|c|}{}&\multicolumn{1}{|c|}{\multirow{2}{*}{${\cal M}_{x\beta}$}}&	\multicolumn{2}{|c|}{\multirow{2}{*}{1.0\%}}&	\multicolumn{2}{|c|}{1.6\%}&	\multicolumn{2}{|c|}{0.9\%}&	\multicolumn{2}{|c|}{0.8\%}\\
			\multicolumn{1}{|c|}{}&	\multicolumn{1}{|c|}{}&	\multicolumn{2}{|c|}{\multirow{2}{*}{}}&	0.30 & 0.50 & 0.22 & 0.45 & 0.21 & 0.45 \\
			\cline{2-10}
			\multicolumn{1}{|c|}{}&\multicolumn{1}{|c|}{\multirow{2}{*}{Fusion}}&		\multicolumn{2}{|c|}{\multirow{2}{*}{0.9\%}}&	\multicolumn{2}{|c|}{1.1\%}&	\multicolumn{2}{|c|}{0.6\%}&	\multicolumn{2}{|c|}{0.5\%}\\
			\multicolumn{1}{|c|}{}&	\multicolumn{1}{|c|}{}&		\multicolumn{2}{|c|}{\multirow{2}{*}{}}&	0.30 & 0.50 &	0.22 & 0.45 &	0.21 & 0.45 \\
			\hline
		\end{tabular}
	\end{center}
	\label{TableMCYT}
\end{table}

\begin{table}[h]
	\caption{\it EERs and BERs for the FVC2000 database. Enrollment methods E1 and E2. }
	\begin{center}
		\begin{tabular}{|c| c|c|c|c c|c c|c c|}
			\hline
			\multicolumn{2}{|c|}{\multirow{2}{*}{\#images ($t$)}}&	\multicolumn{2}{|c|}{\multirow{2}{*}{Analog}}&	\multicolumn{2}{|c|}{\multirow{2}{*}{No HDS}}&	\multicolumn{2}{|c|}{\multirow{2}{*}{ZLHDS}}&	\multicolumn{2}{|c|}{\multirow{2}{*}{ZLHDS+r.c.}}\\
			\multicolumn{2}{|c|}{}&	\multicolumn{2}{|c|}{\multirow{2}{*}{}}&	\multicolumn{2}{|c|}{\multirow{2}{*}{}}&	\multicolumn{2}{|c|}{\multirow{2}{*}{}}&	\multicolumn{2}{|c|}{\multirow{2}{*}{}}\\
			\hline 
			\multicolumn{1}{|c|}{\multirow{6}{*}{1}}&\multicolumn{1}{|c|}{\multirow{2}{*}{${\cal M}_{x\theta}$}}&	\multicolumn{2}{|c|}{\multirow{2}{*}{6.0\% }}&	\multicolumn{2}{|c|}{9.4\%}&	\multicolumn{2}{|c|}{9.0\%}&	\multicolumn{2}{|c|}{8.0\%}\\
			\multicolumn{1}{|c|}{}&	\multicolumn{1}{|c|}{}&	\multicolumn{2}{|c|}{\multirow{2}{*}{}}&0.39  & 0.50  &	0.37 &0.50 &0.36 & 0.50	 \\
			\cline{2-10}
			\multicolumn{1}{|c|}{}&\multicolumn{1}{|c|}{\multirow{2}{*}{${\cal M}_{x\beta}$}}&	\multicolumn{2}{|c|}{\multirow{2}{*}{6.1\%}}&	\multicolumn{2}{|c|}{10.4\%}&	\multicolumn{2}{|c|}{9.5\%}&	\multicolumn{2}{|c|}{8.1\%}\\
			\multicolumn{1}{|c|}{}&	\multicolumn{1}{|c|}{}&	\multicolumn{2}{|c|}{\multirow{2}{*}{}}& 0.39 & 0.50 & 0.38 & 0.50 & 0.37 & 0.50\\
			\cline{2-10}
			\multicolumn{1}{|c|}{}&\multicolumn{1}{|c|}{\multirow{2}{*}{Fusion}}&	\multicolumn{2}{|c|}{\multirow{2}{*}{4.8\%}}&	\multicolumn{2}{|c|}{7.3\%}&	\multicolumn{2}{|c|}{6.5\%}&	\multicolumn{2}{|c|}{5.5\%}\\
			\multicolumn{1}{|c|}{}&	\multicolumn{1}{|c|}{}&		\multicolumn{2}{|c|}{\multirow{2}{*}{}}& 0.39 & 0.50 & 0.38 & 0.50 & 0.36	& 0.50 \\
			\hline 
			\multicolumn{1}{|c|}{\multirow{6}{*}{2}}&\multicolumn{1}{|c|}{\multirow{2}{*}{${\cal M}_{x\theta}$}}&		\multicolumn{2}{|c|}{\multirow{2}{*}{4.5\%}}&	\multicolumn{2}{|c|}{7.2\%}&	\multicolumn{2}{|c|}{5.7\%}&	\multicolumn{2}{|c|}{5.0\%}\\
			\multicolumn{1}{|c|}{}&	\multicolumn{1}{|c|}{}&		\multicolumn{2}{|c|}{\multirow{2}{*}{}}&	0.37 	& 0.50 	& 0.33  & 0.47	& 0.32 & 0.47\\
			\cline{2-10}
			\multicolumn{1}{|c|}{}&\multicolumn{1}{|c|}{\multirow{2}{*}{${\cal M}_{x\beta}$}}&	\multicolumn{2}{|c|}{\multirow{2}{*}{4.8\%}}&	\multicolumn{2}{|c|}{7.9\%}&	\multicolumn{2}{|c|}{6.9\%}&	\multicolumn{2}{|c|}{5.6\%}\\
			\multicolumn{1}{|c|}{}&	\multicolumn{1}{|c|}{}&	\multicolumn{2}{|c|}{\multirow{2}{*}{}}&	0.38 & 0.50 & 0.34 & 0.47 &	0.32	& 0.47\\
			\cline{2-10}
			\multicolumn{1}{|c|}{}&\multicolumn{1}{|c|}{\multirow{2}{*}{Fusion}}&	\multicolumn{2}{|c|}{\multirow{2}{*}{3.9\%}}&	\multicolumn{2}{|c|}{5.1\%}&	\multicolumn{2}{|c|}{5.0\%}&	\multicolumn{2}{|c|}{4.1\%}\\
			\multicolumn{1}{|c|}{}&	\multicolumn{1}{|c|}{}&	\multicolumn{2}{|c|}{\multirow{2}{*}{}}&	0.37 	& 0.50 & 0.33	& 0.47	& 0.32 & 0.47 \\
			\hline 
			\multicolumn{1}{|c|}{\multirow{6}{*}{3}}&\multicolumn{1}{|c|}{\multirow{2}{*}{${\cal M}_{x\theta}$}}&	\multicolumn{2}{|c|}{\multirow{2}{*}{3.0\%}}&	\multicolumn{2}{|c|}{5.6\%}&	\multicolumn{2}{|c|}{5.3\%}&	\multicolumn{2}{|c|}{4.4\%}\\
			\multicolumn{1}{|c|}{}&	\multicolumn{1}{|c|}{}&	\multicolumn{2}{|c|}{\multirow{2}{*}{}}& 0.36 & 0.50 & 0.31	& 0.46 & 0.29 & 0.46 \\
			\cline{2-10}
			\multicolumn{1}{|c|}{}&\multicolumn{1}{|c|}{\multirow{2}{*}{${\cal M}_{x\beta}$}}&	\multicolumn{2}{|c|}{\multirow{2}{*}{3.2\%}}&	\multicolumn{2}{|c|}{7.2\%}&	\multicolumn{2}{|c|}{5.3\%}&	\multicolumn{2}{|c|}{4.9\%}\\
			\multicolumn{1}{|c|}{}&	\multicolumn{1}{|c|}{}&	\multicolumn{2}{|c|}{\multirow{2}{*}{}}&	0.37 & 0.50 & 0.32 & 0.46 & 0.30 & 0.46 \\
			\cline{2-10}
			\multicolumn{1}{|c|}{}&\multicolumn{1}{|c|}{\multirow{2}{*}{Fusion}}&		\multicolumn{2}{|c|}{\multirow{2}{*}{2.2\%}}&	\multicolumn{2}{|c|}{4.5\%}&	\multicolumn{2}{|c|}{4.0\%}&	\multicolumn{2}{|c|}{3.3\%}\\
			\multicolumn{1}{|c|}{}&	\multicolumn{1}{|c|}{}&	\multicolumn{2}{|c|}{\multirow{2}{*}{}}& 0.37 & 0.50 & 0.32 & 0.46 & 0.30 & 0.46\\
			\hline 
			\multicolumn{1}{|c|}{\multirow{6}{*}{4}}&\multicolumn{1}{|c|}{\multirow{2}{*}{${\cal M}_{x\theta}$}}&	\multicolumn{2}{|c|}{\multirow{2}{*}{2.1\%}}&	\multicolumn{2}{|c|}{5.5\%}&	\multicolumn{2}{|c|}{5.5\%}&	\multicolumn{2}{|c|}{4.8\%}\\
			\multicolumn{1}{|c|}{}&	\multicolumn{1}{|c|}{}&	\multicolumn{2}{|c|}{\multirow{2}{*}{}}&	0.37 & 0.50 & 0.31 & 0.45 & 0.29 & 0.45 \\
			\cline{2-10}
			\multicolumn{1}{|c|}{}&\multicolumn{1}{|c|}{\multirow{2}{*}{${\cal M}_{x\beta}$}}&	\multicolumn{2}{|c|}{\multirow{2}{*}{2.2\%}}&	\multicolumn{2}{|c|}{7.1\%}&	\multicolumn{2}{|c|}{6.5\%}&	\multicolumn{2}{|c|}{5.0\%}\\
			\multicolumn{1}{|c|}{}&	\multicolumn{1}{|c|}{}&	\multicolumn{2}{|c|}{\multirow{2}{*}{}}&	0.37 & 0.50 & 0.32 & 0.46 &	0.30 & 0.46 \\
			\cline{2-10}
			\multicolumn{1}{|c|}{}&\multicolumn{1}{|c|}{\multirow{2}{*}{Fusion}}&		\multicolumn{2}{|c|}{\multirow{2}{*}{1.3\%}}&	\multicolumn{2}{|c|}{4.3\%}&	\multicolumn{2}{|c|}{4.3\%}&	\multicolumn{2}{|c|}{3.3\%}\\
			\multicolumn{1}{|c|}{}&	\multicolumn{1}{|c|}{}&		\multicolumn{2}{|c|}{\multirow{2}{*}{}}&	0.37 & 0.50 &	0.31 & 0.45 & 0.30 & 0.45\\
			\hline
		\end{tabular}
	\end{center}
	\label{TableFVC2000}
\end{table}

\begin{table}[h]
	\caption{\it EERs and BERs for the FVC2002 database. Enrollment methods E1 and E2.}
	\begin{center}
		\begin{tabular}{|c| c|c|c|cc|cc|cc|}
			\hline
			\multicolumn{2}{|c|}{\multirow{2}{*}{\#images ($t$)}}&	\multicolumn{2}{|c|}{\multirow{2}{*}{Analog}}&	\multicolumn{2}{|c|}{\multirow{2}{*}{No HDS}}&	\multicolumn{2}{|c|}{\multirow{2}{*}{ZLHDS}}&	\multicolumn{2}{|c|}{\multirow{2}{*}{ZLHDS+r.c.}}\\
			\multicolumn{2}{|c|}{}&	\multicolumn{2}{|c|}{\multirow{2}{*}{}}&	\multicolumn{2}{|c|}{\multirow{2}{*}{}}&	\multicolumn{2}{|c|}{\multirow{2}{*}{}}&	\multicolumn{2}{|c|}{\multirow{2}{*}{}}\\
			\hline 
			\multicolumn{1}{|c|}{\multirow{6}{*}{1}}&\multicolumn{1}{|c|}{\multirow{2}{*}{${\cal M}_{x\theta}$}}&	\multicolumn{2}{|c|}{\multirow{2}{*}{5.8\%}}&	\multicolumn{2}{|c|}{12.1\%}&	\multicolumn{2}{|c|}{10.8\%}&	\multicolumn{2}{|c|}{8.8\%}\\
			\multicolumn{1}{|c|}{}&	\multicolumn{1}{|c|}{}&	\multicolumn{2}{|c|}{\multirow{2}{*}{}}& 0.38 & 0.50 & 0.37 & 0.50 & 0.36 & 0.50 \\
			\cline{2-10}
			\multicolumn{1}{|c|}{}&\multicolumn{1}{|c|}{\multirow{2}{*}{${\cal M}_{x\beta}$}}&	\multicolumn{2}{|c|}{\multirow{2}{*}{6.4\%}}&	\multicolumn{2}{|c|}{10.9\%}&	\multicolumn{2}{|c|}{10.9\%}&	\multicolumn{2}{|c|}{9.2\%}\\
			\multicolumn{1}{|c|}{}&	\multicolumn{1}{|c|}{}&	\multicolumn{2}{|c|}{\multirow{2}{*}{}}&	0.39 & 0.50 & 0.38 & 0.50 &	0.36 & 0.50	\\
			\cline{2-10}
			\multicolumn{1}{|c|}{}&\multicolumn{1}{|c|}{\multirow{2}{*}{Fusion}}&	\multicolumn{2}{|c|}{\multirow{2}{*}{5.5\%}}&	\multicolumn{2}{|c|}{9.4\%}&	\multicolumn{2}{|c|}{9.3\%}&	\multicolumn{2}{|c|}{7.0\%}\\
			\multicolumn{1}{|c|}{}&	\multicolumn{1}{|c|}{}&		\multicolumn{2}{|c|}{\multirow{2}{*}{}}& 0.39 & 0.50 & 0.38 & 0.50 & 0.36 & 0.50 \\
			\hline 
			\multicolumn{1}{|c|}{\multirow{6}{*}{2}}&\multicolumn{1}{|c|}{\multirow{2}{*}{${\cal M}_{x\theta}$}}&		\multicolumn{2}{|c|}{\multirow{2}{*}{5.4\%}}&	\multicolumn{2}{|c|}{10.9\%}&	\multicolumn{2}{|c|}{9.8\%}&	\multicolumn{2}{|c|}{7.3\%}\\
		\multicolumn{1}{|c|}{}&	\multicolumn{1}{|c|}{}&		\multicolumn{2}{|c|}{\multirow{2}{*}{}}& 0.39 & 0.50 & 0.35 & 0.48 & 0.33 & 0.48 \\
		\cline{2-10}
		\multicolumn{1}{|c|}{}&\multicolumn{1}{|c|}{\multirow{2}{*}{${\cal M}_{x\beta}$}}&	\multicolumn{2}{|c|}{\multirow{2}{*}{5.5\%}}&	\multicolumn{2}{|c|}{10.7\%}&	\multicolumn{2}{|c|}{8.4\%}&	\multicolumn{2}{|c|}{7.4\%}\\
		\multicolumn{1}{|c|}{}&	\multicolumn{1}{|c|}{}&	\multicolumn{2}{|c|}{\multirow{2}{*}{}}& 0.39 & 0.50 & 0.36 & 0.48 & 0.34 & 0.48 \\
		\cline{2-10}
		\multicolumn{1}{|c|}{}&\multicolumn{1}{|c|}{\multirow{2}{*}{Fusion}}&	\multicolumn{2}{|c|}{\multirow{2}{*}{4.4\%}}&	\multicolumn{2}{|c|}{9.8\%}&	\multicolumn{2}{|c|}{7.3\%}&	\multicolumn{2}{|c|}{5.9\%}\\
		\multicolumn{1}{|c|}{}&	\multicolumn{1}{|c|}{}&	\multicolumn{2}{|c|}{\multirow{2}{*}{}}& 0.39 & 0.50 & 0.36 & 0.48 & 0.34 & 0.48\\
		
		\hline
	\end{tabular}
\end{center}
\label{TableFVC2002}
\end{table}

\begin{table}[h]
	\caption{\it EERs and BERs for the FVC2000 database. Enrollment method E3.}
	\begin{center}
		\begin{tabular}{|c|c|c|c|cc|cc|cc|}
			\hline
			\multicolumn{2}{|c|}{\multirow{2}{*}{\#images ($t$)}}&	\multicolumn{2}{|c|}{\multirow{2}{*}{Analog}}&	\multicolumn{2}{|c|}{\multirow{2}{*}{No HDS}}&	\multicolumn{2}{|c|}{\multirow{2}{*}{ZLHDS}}&	\multicolumn{2}{|c|}{\multirow{2}{*}{ZLHDS+r.c.}}\\
			\multicolumn{2}{|c|}{}&	\multicolumn{2}{|c|}{\multirow{2}{*}{}}&	\multicolumn{2}{|c|}{\multirow{2}{*}{}}&	\multicolumn{2}{|c|}{\multirow{2}{*}{}}&	\multicolumn{2}{|c|}{\multirow{2}{*}{}}\\
			\hline 
			\multicolumn{1}{|c|}{\multirow{6}{*}{3}}&\multicolumn{1}{|c|}{\multirow{2}{*}{${\cal M}_{x\theta}$}}&	\multicolumn{2}{|c|}{\multirow{2}{*}{3.0\%}}&	\multicolumn{2}{|c|}{5.8\%}&	\multicolumn{2}{|c|}{5.2\%}&	\multicolumn{2}{|c|}{4.2\%}\\
			\multicolumn{1}{|c|}{}&	\multicolumn{1}{|c|}{}&	\multicolumn{2}{|c|}{\multirow{2}{*}{}}& 0.37 & 0.50  & 0.36 & 0.50 & 0.34 & 0.50 \\
			\cline{2-10}
			\multicolumn{1}{|c|}{}&\multicolumn{1}{|c|}{\multirow{2}{*}{${\cal M}_{x\beta}$}}&	\multicolumn{2}{|c|}{\multirow{2}{*}{3.2\%}}&	\multicolumn{2}{|c|}{8.1\%}&	\multicolumn{2}{|c|}{6.1\%}&	\multicolumn{2}{|c|}{5.4\%}\\
			\multicolumn{1}{|c|}{}&	\multicolumn{1}{|c|}{}&	\multicolumn{2}{|c|}{\multirow{2}{*}{}}& 0.37 & 0.50 & 0.36 & 0.50 & 0.35 & 0.50 \\
			\cline{2-10}
			\multicolumn{1}{|c|}{}&\multicolumn{1}{|c|}{\multirow{2}{*}{Fusion}}&		\multicolumn{2}{|c|}{\multirow{2}{*}{2.2\%}}&	\multicolumn{2}{|c|}{5.3\%}&	\multicolumn{2}{|c|}{4.0\%}&	\multicolumn{2}{|c|}{3.1\%}\\
			\multicolumn{1}{|c|}{}&	\multicolumn{1}{|c|}{}&	\multicolumn{2}{|c|}{\multirow{2}{*}{}}& 0.37 & 0.50 & 0.36 & 0.50 & 0.34 & 0.50	\\
			\hline
		\end{tabular}
	\end{center}
	\label{TableFVC2000MV}
\end{table}

\begin{table}[h]
	\caption{\it EERs and BERs for the MCYT database. Enrollment method E3.}
	\begin{center}
		\begin{tabular}{|c| c|c|c|cc|cc|cc|}
			\hline
			\multicolumn{2}{|c|}{\multirow{2}{*}{\#images ($t$)}}&	\multicolumn{2}{|c|}{\multirow{2}{*}{Analog}}&	\multicolumn{2}{|c|}{\multirow{2}{*}{No HDS}}&	\multicolumn{2}{|c|}{\multirow{2}{*}{ZLHDS}}&	\multicolumn{2}{|c|}{\multirow{2}{*}{ZLHDS+r.c.}}\\
			\multicolumn{2}{|c|}{}&	\multicolumn{2}{|c|}{\multirow{2}{*}{}}&	\multicolumn{2}{|c|}{\multirow{2}{*}{}}&	\multicolumn{2}{|c|}{\multirow{2}{*}{}}&	\multicolumn{2}{|c|}{\multirow{2}{*}{}}\\
			\hline 
			\multicolumn{1}{|c|}{\multirow{6}{*}{3}}&\multicolumn{1}{|c|}{\multirow{2}{*}{${\cal M}_{x\theta}$}}&	\multicolumn{2}{|c|}{\multirow{2}{*}{1.4\%}}&	\multicolumn{2}{|c|}{2.4\%}&	\multicolumn{2}{|c|}{1.6\%}&	\multicolumn{2}{|c|}{1.4\%}\\
			\multicolumn{1}{|c|}{}&	\multicolumn{1}{|c|}{}&	\multicolumn{2}{|c|}{\multirow{2}{*}{}}& 0.31 & 0.50 & 0.29 & 0.49 & 0.28 & 0.49\\
			\cline{2-10}
			\multicolumn{1}{|c|}{}&\multicolumn{1}{|c|}{\multirow{2}{*}{${\cal M}_{x\beta}$}}&	\multicolumn{2}{|c|}{\multirow{2}{*}{1.1\%}}&	\multicolumn{2}{|c|}{2.2\%}&	\multicolumn{2}{|c|}{1.5\%}&	\multicolumn{2}{|c|}{1.4\%}\\
			\multicolumn{1}{|c|}{}&	\multicolumn{1}{|c|}{}&	\multicolumn{2}{|c|}{\multirow{2}{*}{}}& 0.32 & 0.50 & 0.30 & 0.50 & 0.28 & 0.50\\
			\cline{2-10}
			\multicolumn{1}{|c|}{}&\multicolumn{1}{|c|}{\multirow{2}{*}{Fusion}}&		\multicolumn{2}{|c|}{\multirow{2}{*}{1.1\%}}&	\multicolumn{2}{|c|}{1.6\%}&	\multicolumn{2}{|c|}{1.0\%}&	\multicolumn{2}{|c|}{0.9\%}\\
			\multicolumn{1}{|c|}{}&	\multicolumn{1}{|c|}{}&	\multicolumn{2}{|c|}{\multirow{2}{*}{}}& 0.32 & 0.50 & 0.30 & 0.49 & 0.28 & 0.50 \\
			\hline
		\end{tabular}
	\end{center}
	\label{TableMCYTMV}
\end{table}


\subsection{Error correction: Polar codes}
\label{sec:polar}

The error rates in the genuine reconstructed $\hat k$ are rather high,
at least 0.21.
In order to apply the Code Offset Method with a decent message size
it is necessary to use a code that has a high rate even at small
codeword length.

Consider the case of fusion of ${\cal M}_{x\theta}$ and ${\cal M}_{x\beta}$.
The codeword length is 1280 bits (1024 if reliable component selection is performed).
Suppose we need to distinguish between $2^{20}$ users. Then the message length needs to
be at least 20 bits, in spite of the high bit error rate.
Furthermore, the security of the template protection is determined by the entropy
of the data that is input into the hash function
(see Fig.\,\ref{fig:twostageHDS}); it would be preferable to have
at least 64 bits of entropy.

We constructed a number of Polar codes tuned to the 
signal-to-noise ratios of the individual grid points. 
The codes are designed to find a set of reliable channels,
which are then assigned to the information bits.
Each code yields a certain FAR 
(impostor string accidentally decoding correctly)
and FRR (genuine reconstruction string failing to decode correctly), 
and hence can be represented as a point in an ROC plot.
This is shown in Fig.\,\ref{fig:codes}.
For the MCYT database we have constructed a Polar code with message length 25
at an EER around 1.2\%
(compared to 0.7\% before error correction).
For the FVC2000 database we have constructed a Polar code with message length 15
at an EER around 6\%
(compared to 3.3\% EER before error correction).
Note that the error correction is an indispensable part of the privacy protection
and inevitably leads to a performance penalty.
However, we see that the penalty is not that bad, especially for high-quality fingerprints.

From our results we also see that even under the best circumstances
(high-quality MCYT database) 
the entropy of the extracted string is severely limited ($\leq$25 bits).
In order to achieve a reasonable security level of the hash,
at least two fingers need to be combined.
We do not see this as a drawback of our helper data system;
given that the EER for one finger is around 1\%, 
which is impractical in real-life applications,
{\it it is necessary anyhow to combine multiple fingers}.

\subsection{Error correction: random codebooks}
There is a large discrepancy between the message length of the Polar code
($k\leq 25$) and the known information content of a fingerprint.
According to Ratha et al \cite{RCB2001} the reproducible entropy of a fingerprint image
with $Z=35$ robustly detectable minutiae should be more than 120 bits.
Furthermore, the potential message size that can be carried 
in a 1024-bit string with a BER of 23\% is $1024[1-h(0.23)]=227$ bits.
(And 122 bits at 30\% BER.)

\begin{figure*}
\begin{center}
\includegraphics[width=0.7\textwidth]{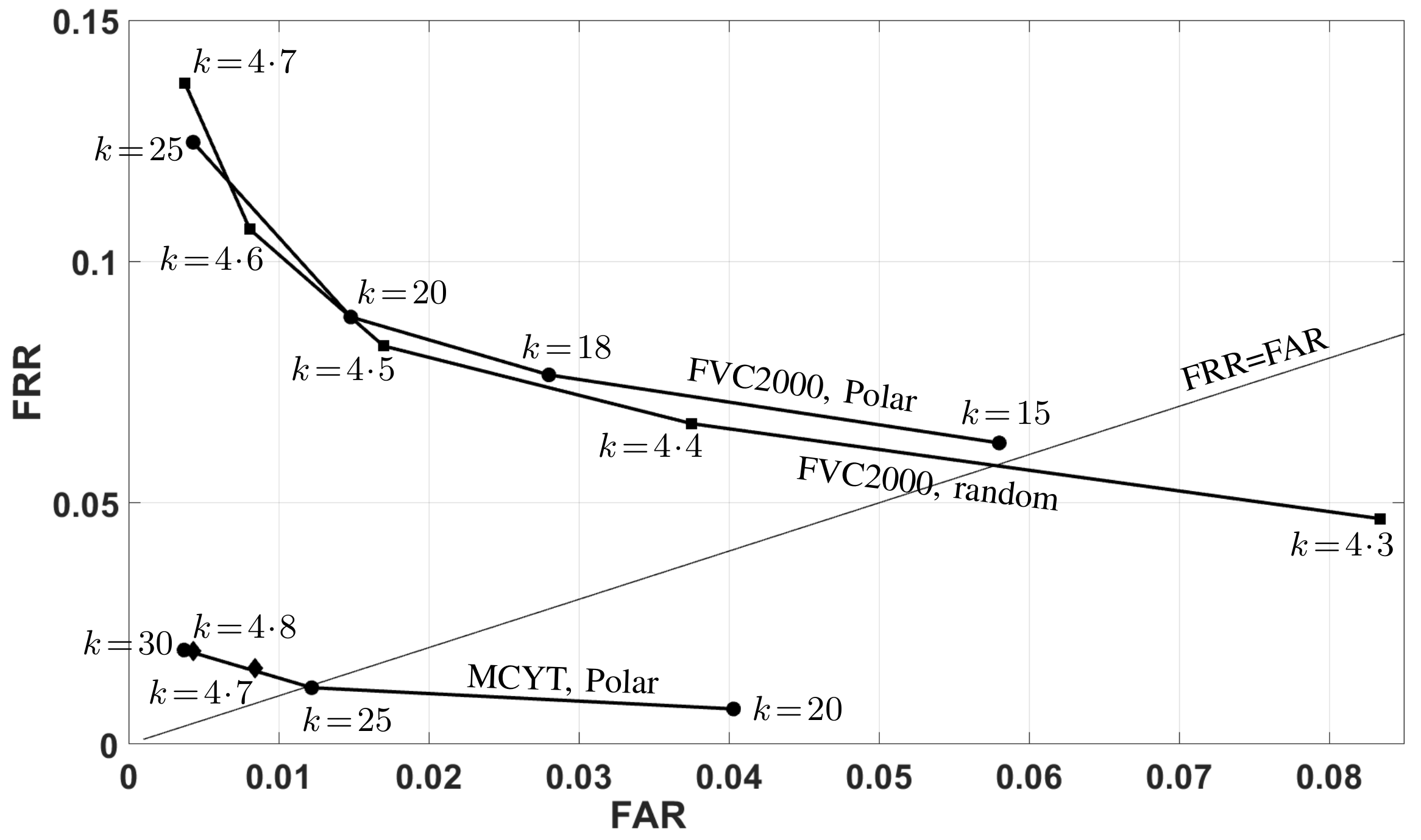}
\end{center}
\caption{ \it
FRR versus FAR achieved by Polar codes and random codebooks (average over three random codebooks).
The $\blacklozenge$ markers denote random codebook, and happen to coincide with the line
connecting the Polar markers.
}
\label{fig:codes}
\end{figure*}

We experimented with random codebooks to see if we could extract more entropy from the data 
than with polar codes.
At low code rates, a code based on random codewords can be practical to implement.
Let the message size be $\ell$, and the codeword size~$m$.
A random table needs to be stored of size $2^\ell\cdot m$ bits, and the process of decoding
consists of computing $2^\ell$ Hamming distances. 
We split the 1024 reliable bits into 4 groups of $m=256$ bits,
for which we generated random codebooks, for various values of~$\ell$. 
The total message size is $k=4\ell$ and the total codeword size is~$n=4m$.
The results are shown in Fig.\,\ref{fig:codes}.
In short: random codebooks give hardly any improvement over Polar codes.

\section{Summary and discussion}
\label{sec:discussion}

A Helper Data System protects privacy but causes a fingerprint recognition degradation in the
form of increased EER.
We have built a HDS from a spectral function representation of fingerprint data,
combined with a Zero Leakage quantisation scheme.
It turns out that our HDS causes only a very small EER penalty when the fingerprint quality is high.

The best results were obtained with the `superfinger' enrollment method 
(E2, taking the average over multiple enrollment images in the spectral function domain),
and with fusion of the ${\cal M}_{x\theta}$,${\cal M}_{x\beta}$ functions.
The superfinger method performs slightly better than the E3 method and also has the advantage
that it is not restricted to an odd number of enrollment captures.

For the high-quality MCYT database, our HDS achieves an EER around 1\% and extracts a 1024-bit string
with $\leq25$ bits of entropy.
In practice multiple fingers need to be used in order to obtain an acceptable EER.
This automatically increases the entropy of the hashed data.
The entropy can be further increased by employing tricks like the
Spammed Code Offset Method \cite{SdV2014}. 

As topics for future work we mention 
(i) testing the HDS on more databases; 
(ii) further optimisation of parameter choices such as the number of reliable components, 
and the number of minutiae used in the computation of the spectral functions;
(iii) further tweaking of the Polar codes.



\section*{Acknowledgments}
Part of this work was supported by NWO grant 628.001.019 (ESPRESSO),
and grant 61701155 from the National Natural Science Foundation of China (NSFC).

\ifCLASSOPTIONcaptionsoff
  \newpage
\fi



%


\bibliographystyle{plain}
\bibliography{spectralSS}

%








\end{document}